\documentclass[structabstract]{aa}
\usepackage{graphicx}
\usepackage{natbib}
\usepackage[dvipsnames]{xcolor}
\usepackage{hyperref}
\usepackage{txfonts}
\usepackage{mathrsfs}
\usepackage{longtable}
\usepackage{tipa}

\newcommand{\ubv}{UBV}
\newcommand{\ubvr}{UBVR}

\newcommand{\hp}{\hbox{$H_{\rm p}$}}

\title{Configuration of the $\xi$~Tau system constrained by multi-technique observations}

\titlerunning{Configuration of the $\xi$~Tau system}

\author{
  M.~Bro\v{z}\inst{\ref{prague}}         \and 
  P.~Dole\v{z}al\inst{\ref{prague}}      \and 
  D.~Vokrouhlick\'{y}\inst{\ref{prague}} \and 
  P.~Harmanec\inst{\ref{prague}}         \and 
  B.~Barlow\inst{\ref{highpoint}}        \and 
  H.~Bo\v{z}i\'c\inst{\ref{hvar}}        \and 
  J.~Labadie-Bartz\inst{\ref{meudon},\ref{lyngby}}    \and 
  R.~Kuschnig\inst{\ref{graz}}           \and 
  T.~Kallinger\inst{\ref{vienna}}        \and 
  J.~Matthews\inst{\ref{vancouver}}      \and 
  D.~Mourard\inst{\ref{nice}}                 
}

\institute{
  Charles University, Faculty of Mathematics and Physics, Institute of Astronomy, V~Hole{\v s}ovi{\v c}k{\'a}ch 2, CZ-18000 Prague, Czech Republic\label{prague}\and
  High Point University, Department of Physics, One University Way, High Point, NC 27268, USA\label{highpoint}\and
  University of Zagreb, Faculty of Geodesy, Hvar Observatory, Ka\v{c}i\'ceva~26, 10000~Zagreb, Croatia\label{hvar}\and
  Observatoire de Paris, LIRA, Université PSL, CNRS, Sorbonne Université, Université Paris Cité, CY Cergy Paris Université, 92190 Meudon, France\label{meudon}\and
  Technical University of Denmark, DTU Space, Elektrovej 327, Kgs., Lyngby, 2800, Denmark\label{lyngby}\and
  Universit\"at Graz, Institut für Physik, Universit\"atsplatz 3, 8010 Graz, Austria\label{graz}\and
  Universit\"at Wien, Institut für Astrophysik, Türkenschanzstrasse 17, 1180 Vienna, Austria\label{vienna}\and
  University of British Columbia, Department of Physics and Astronomy, 6224 Agricultural Road, Vancouver, BC V6T 1Z1, Canada\label{vancouver}\and
  Universit\'e C\^ote d'Azur, Observatoire de la C\^ote d'Azur, CNRS, Laboratoire Lagrange, Parc Valrose, 06108 Nice, France\label{nice}
}

\date{Received x-x-2026 / Accepted x-x-2026}

\abstract
   {
$\xi$ Tau is one of the most compact multiple stellar systems,
which is sufficiently close (67\,pc)
to be constrained by all kinds of observations.
   }
   {
To better constrain the current configuration of $\xi$~Tau
and to study its temporal evolution,
we utilized new observational data:
(i)~publicly available photometry from TESS
and astrometry from WDS,
and (ii)~our own photometry from the MOST spacecraft
and spectroscopy from the CTIO observatory.
   }
   {
As a prerequisite,
we accurately described the CTIO/CHIRON echelle instrument
so that interpolation of the blaze function
over the Balmer lines (H$\alpha$, H$\beta$)
allowed us to rectify the wings
with minimal systematics.
We then used our previous model of $\xi$~Tau
and constrained it by the new data.
At the same time, we improved our dynamical model by accounting
not only for mutual, N-body perturbations, but also for
relativistic effects,
oblateness, and
tides.
We optimized grids of models,
showing $\chi^2$ contributions from 11~different types of observations
(astrometry,
radial velocities,
transit-timing variations,
light curves,
interferometric visibility,
closure phase,
\dots).
This allowed us to find the global, best-fit solution,
keeping track of possible `tension' among the different data.
   }
   {
Given the hierarchical architecture of $\xi$~Tau,
((Aa+Ab)+B)+C,
we detected the orbital evolution on all time scales.
Oscillations of periods $P$ occur on the shortest, orbital time scales
($P_1$, $P_2$);
the variation of eccentricity $e_1$ is from 0 to 0.008,
and of $e_2$ from 0.202 to 0.207,
respectively.
Oscillations of projected $i$, $\Omega$ are coupled,
and occur on the secular time scale of about
7000\,d.
The inclination $i_1$ of the inner, eclipsing pair (Aa+Ab)
changes from $86.1^\circ$ to $87.1^\circ$,
which is clearly manifested in eclipse depths.
There is also a long-term trend due to the outer orbit
($P_3 \doteq 18900\,{\rm d}$),
with a perihelion passage (a `bump') of component~C,
which is manifested in radial velocities.
The mutual inclinations between the three orbital planes,
${\simeq}\,0.5^\circ$ and $71^\circ$,
are very different.
Long-term stability is ensured by suppressing Kozai oscillations
due to the fast precession rate $\dot\omega_2$.
The best model requires tidal dissipation in the inner binary
(with the time lag of ${\sim}100\,{\rm s}$)
and five components, where component C is a binary (Ca+Cb).
   }
   {
Although the masses of the three components ($2.27$, $2.15$, $3.78\,M_\odot$)
are now constrained to within 1\,\%,
the suspected binary (Ca+Cb), offset by $600\,{\rm mas}$,
should be better characterized.
A key question remains whether this bright stellar system
contains additional dwarf or exoplanetary components with low masses.
Continuing monitoring of $\xi$~Tau is highly desirable.
   }


\begin{document}

\maketitle


\section{Introduction}

Triple, quadruple, or quintuple stellar systems are not just
'more complicated binaries'
\citep{Tokovinin_2020MNRAS.491.5158T,Tokovinin_2021Univ....7..352T,Borkovits_2020MNRAS.496.4624B,Borkovits_2022Galax..10....9B}.
They were not assembled randomly,
but their varying architectures and occurrences
(2+1, 3+1, 2+2, 3+2, \dots)
suggest several evolutionary processes, {\em par exemple}\footnote{to be pronounced with a French-Belgian accent},
migration,
Kozai cycles, 
tidal friction,
magnetic friction,
envelope expansion,
mass transfer,
close encounters, or
ejections.

Less complicated binaries form a foundation of the universal distance scale
\citep{Pietrzynski_2013Natur.495...76P,Pietrzynski_2019Natur.567..200P,Gallenne_2023A&A...672A.119G},
provided their properties were measured with per-cent precision.
Multiples allow to achieve even higher precision
due to additional dynamical constraints
\citep{Borkovits_2022Galax..10....9B,Broz_2025arXiv250620866B}.
Two or more orbits might serve also as a double-check.

The $\xi$~Tau system is composed of four components
(Aa, Ab, B, C).
From a solar-system perspective,
the inner eclipsing binary has a semimajor axis of only $0.11\,{\rm au}$,
the 3rd component like the Earth (${\sim}1.1\,{\rm au}$),
the 4th component like Neptune ($29\,{\rm au}$),
with a total mass of more than $9\,M_\odot$.

In our previous work
\citep{Nemravova_2016A.A...594A..55N},
we used space-based photometry from
the Microvariability and Oscillations of Stars Satellite (MOST)
\citep{Walker_2003PASP..115.1023W},
which allowed a clear detection of mutual perturbations;
for example,
the eclipse-timing variations (ETVs)
exhibit an amplitude of about half an hour.
In this work,
we aim at using space-based photometry from the Transiting Exoplanet Survey Satellite (TESS)
\citep{tess} and additional set of MOST observations from 2017,
supplemented by ground-based spectroscopy from Cerro Tololo Inter-American Observatory (CTIO)
\citep{Tokovinin_2013PASP..125.1336T}.
Consequently, the high-accuracy observations extend over a substantially longer time span.
Their analysis should allow a much better characterization of the system parameters,
including evidence of the evolution of $\xi$~Tau orbital architecture on various timescales
(from days up to decades).

A related scientific question is
whether or not this stellar system also contain exoplanets.
More than 800 exoplanets are already known in binaries
\citep{Thebault_2015pes..book..309T,Thebault_2025A&A...700A.106T}.
More than 30 are circum-binary,
where a secondary is commonly located at $0.1\,{\rm au}$
and an exoplanet at ${\sim}0.5\,{\rm au}$,
which is close to the stability limit
\citep{Holman_1999AJ....117..621H}.
This rough statistics is biased though,
especially for O or B stars,
which are much brighter than exoplanets.
One should keep it in mind when studying bright stellar systems.
\vskip\baselineskip


Hereinafter, we describe only new observations of $\xi$~Tau
(Sect.~\ref{observations}),
since old observations were described in \citet{Nemravova_2016A.A...594A..55N}.
Subsequently, we describe models of $\xi$~Tau of various complexity,
the reference one (Sect.~\ref{3.1}),
the interferometric one (Sect.~\ref{3.2}),
the all-data model (Sect.~\ref{3.3}),
the model with tides (Sect.~\ref{tides}), or
the model with five components (Sect.~\ref{five}).
The temporal evolution of the $\xi$~Tau system is discussed in Sect.~\ref{evolution}
and additional aspects (parallax, oscillations)
in Sect.~\ref{discussion}.
The conclusions are presented in Sect.~\ref{conclusions}.


\section{New observations}\label{observations}

\subsection{Hvar photometry}\label{photometric}

Systematic photometry was collected at the Hvar Observatory,
with the 0.65-m reflector and photoelectric photometer;
initially in UBV bands (2007--2013)
and later in UBVR bands (2013--2025).
After a recent revision of 52 years of Hvar observations
\citep{Bozic_2026},
all-sky magnitudes of all comparison stars were derived
relative to the Johnson standards
and all observations were carefully transformed to the standard Johnson UBV system
\citep{hhj94}.
The latest reduction includes also
temporal variation of the extinction coefficients
in the course of night.
The Hvar mean values for comparison stars, 4~Tau = HD~21686, and 6~Tau = HD~21933, 
which were added to the differential magnitudes, are
$V=5.150$, $B-V = -0.045$, $U-B = -0.093$, $V-R = -0.016$;
$V=5.774$, $B-V = -0.076$, $U-B = -0.300$, $V-R = -0.025$.

We note that the transparency curve of the R filter
closely corresponds to that of the standard Cousins R filter.
However, because we were not able to find enough northern bright standard stars
with the Cousins $R_{\rm c}$ values,
we derived robust mean values of Johnson $V-R$ indices from \citet{john66}
and reduced our observations to the Johnson $R$ magnitude.






\subsection{MOST photometry}

Broadband photometry obtained by the MOST satellite
\citep{Walker_2003PASP..115.1023W}
consists of two separate sets of observations.
Back in 2012, J.~Nemravov\'a and P.~Harmanec submitted a successful application
for observations of $\xi$~Tau to the MOST allocation committee,
and continuous light curve observations were indeed obtained.
These observations have been published in \citet{Nemravova_2016A.A...594A..55N},
along with the discovery of low-amplitude photometric oscillations
on the time scale of about 0.4\,d. We use these data also here.

The second data set was obtained on a commercial basis at the end of 2017
and represents the last set of scientifically usable MOST data,
before the communication with the satellite was lost forever.
The data were downloaded at the Vienna tracking station
and their initial reduction was carried out by T.~Kallinger.


\subsection{TESS photometry}

Photometry obtained by the TESS satellite
\citep{tess}
was also included.
TESS records red optical light with a wide bandpass
spanning roughly 600--1000\,nm,
centered on the traditional Cousins $I$ band.
TESS observed $\xi$~Tau in six sectors (31, 42, 43, 44, 70, 71).
Basic reductions and cleaning were performed by Jon Labadie-Bartz.
Light curves were extracted from the full frame images with
the Lightkurve package
\citep{Cardoso_2018ascl.soft12013L},
using simple aperture photometry,
including the saturated columns and their ``spillover'' pixel end caps,
and the non-saturated halo pixels,
as is standard for moderately saturated bright stars.
The observing cadence was 10 minutes (sectors 31, 42, 43, 44)
and 200 seconds (sectors 70, 71).
Background subtraction was used as the preferred detrending method.
$\xi$~Tau is fairly isolated on the sky
and so blending from neighboring stars
did not contribute to the observed signals.

The series of light curves represents the key observational constraint,
because the exact epochs of eclipses are sensitive to the perturbations
by the component~B.
These include both the geometric effects due to finite speed of light
and variations in position of the eclipsed star with respect to the observer
(usually termed light-time effect),
and variations in the orbital elements of the eclipsing binary
due to the gravitational perturbations by the component~B
(often termed physical timing variations).


\subsection{CTIO/CHIRON spectroscopy}\label{spectroscopic}

The CHIRON instrument
\citep{Tokovinin_2013PASP..125.1336T}
at the CTIO 1.5-m telescope
is an echelle spectrograph.
Its resolution is $R = 79000$ (higher resolution was not used),
spectral range 410 to 890\,nm,
efficiency at least 6\%.
It is equipped with the back-illuminated detector CCD231-84 Teledyne E2v,
with $4094\,{\times}\,4112$~pixels,
and gain $\eta = 1.3\,{\rm adu}/{\rm e}^-$.
We performed standard reduction of spectra, by applying
dark frame,
flat field,
and by calibrating the wavelength scale
with a ThAr lamp spectrum.
For the detailed description of rectification,
see Appendix~\ref{rectification}.

In the course of observational campaign,
we obtained 277 spectra of $\xi$~Tau
(excluding defective ones).
The spectral range was limited to 450 to 890\,nm,
divided into 62 orders
(including some overlaps and gaps),
with 801 pixels per order.
The exposure time was 60\,s,
the peak signal reaching $S = 160000\,{\rm adu}$,
resulting in the signal-to-noise ratio $S/N = \sqrt{S/\eta} = 350$.
We observed sequences of 5 spectra,
the total time span was 129~days,
from Sep 15th 2021 to Jan 21st 2022,
covering the short orbit (Aa+Ab) 18~times,
and about 90 \% of the long orbit ((Aa+Ab)+B).

\begin{figure*}
\centering
\includegraphics[width=13cm]{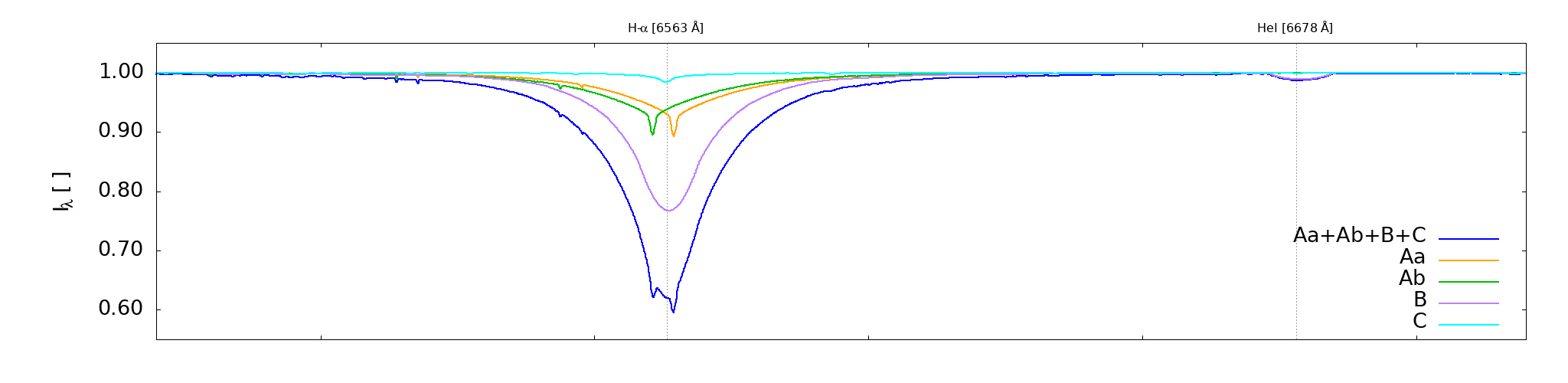}
\includegraphics[width=13cm]{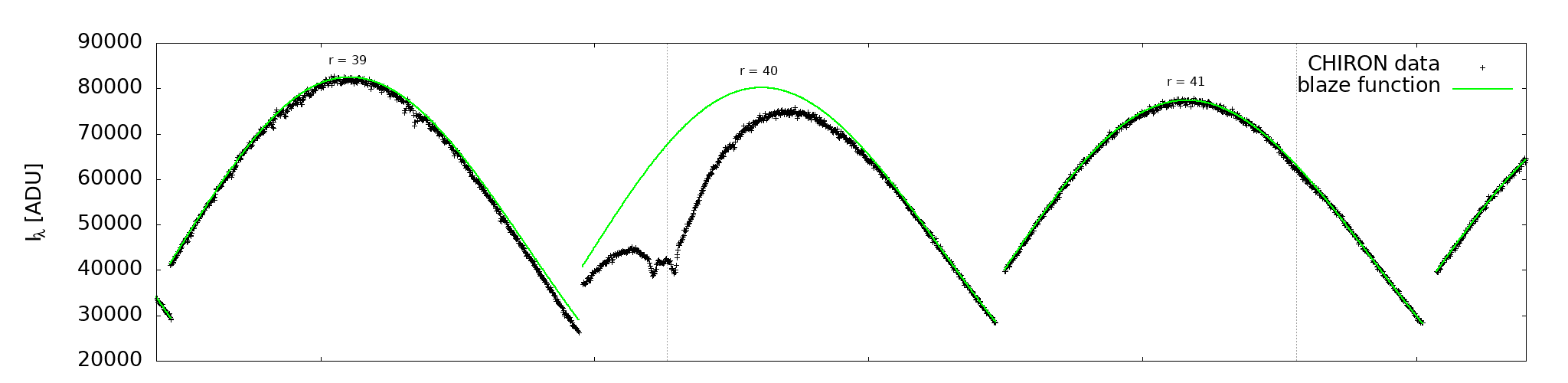}
\includegraphics[width=13cm]{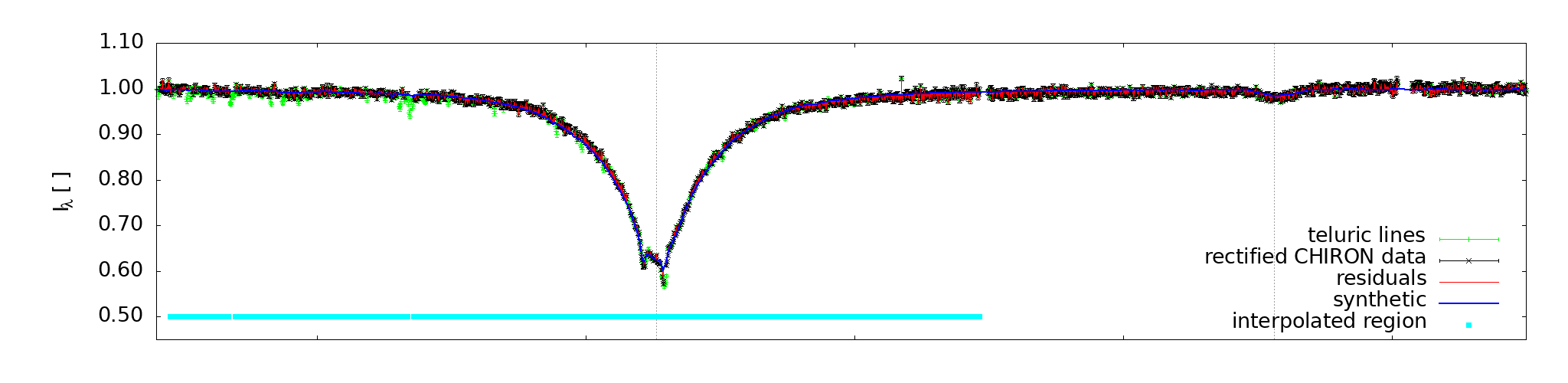}
\includegraphics[width=13cm]{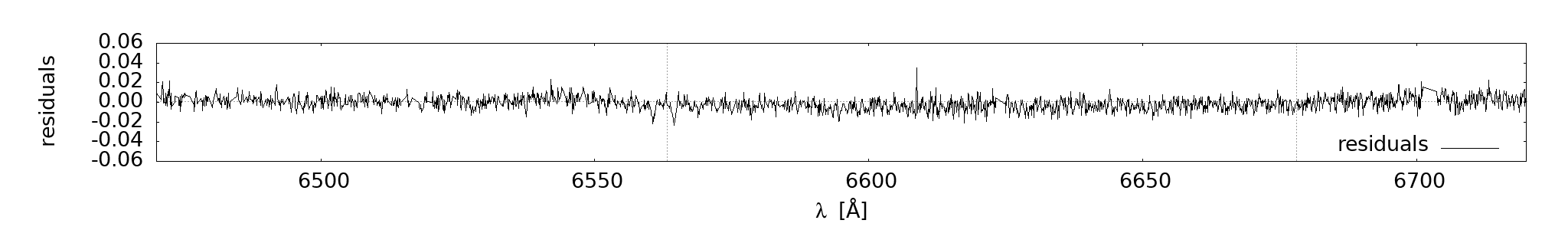}
\caption{
Rectification of a CTIO/CHIRON echelle spectrum (ktc00027)
in the region of the H$\alpha$ line (6563\,\AA),
using the resimplex3 method.
Top:
Synthetic spectra were computed from our model in Table~\ref{params}, column "all-data".
Four component spectra (Aa, Ab, B, C) and the resulting spectrum are plotted.
2nd row:
Observed spectrum from CHIRON (black),
orders $r = 39$ to $41$,
and the corresponding blaze function (\color{green}green\color{black}).
3rd row:
Rectified spectrum (black),
its comparison to the synthetic spectrum (\color{blue}blue\color{black})
and corresponding residuals (\color{red}red\color{black}).
The rectification was done with respect to the continuum level,
with interpolation of the blaze parameters over wide lines (\color{cyan}cyan\color{black})
and problematic orders,
avoiding also locations of the telluric lines (\color{green}green\color{black}).
Bottom:
Residuals plotted separately.
}
\label{synthetic_comps}
\end{figure*}


\subsection{Gaia parallax}

The parallax in the Gaia EDR3/DR3 release
\citep{Bailer_2021AJ....161..147B,Vallenari_2023A&A...674A...1G}
is
$\pi = (16.8\pm 0.7)\,{\rm mas}$,
equivalent to
$d = (59.6\pm 2.4)\,{\rm pc}$.
On the contrary, the previous modeling of
\citet{Nemravova_2016A.A...594A..55N}
lead to the different distance
$d = (67.9\pm 1.0)\,{\rm pc}$,
which is likely the correct one,
because $\xi$~Tau exhibits photocentre motion,
negatively impacting the parallactic measurement.



\subsection{WDS astrometry}

Seven new astrometric measurements from the Washington Double Star (WDS) catalogue
\citep{Mason_2001AJ....122.3466M,Tokovinin_2020AJ....160....7T}
were included.
Together with a removal of two outlier points (in Julian date, TDB)
2456314.597291,
2456936.276038,
the new datset provides a slightly different outer orbit solution,
compared to the previous modeling in \cite{Nemravova_2016A.A...594A..55N}.



\subsection{Other data}

Description of other data was presented in
\cite{Nemravova_2016A.A...594A..55N},
namely of
old radial velocities (RV),
eclipse timing variations (ETV),
eclipse durations,
interferometric visibility,
closure phase,
triple product, and
the spectral-energy distribution (SED).
Of course, these datasets were used also in this study,
in order to fully constrain the model of $\xi$~Tau.



\section{Configuration of $\xi$~Tau} \label{configuration}

To describe the dynamics of $\xi$~Tau,
we use the eponymous model
{\tt Xitau}\footnote{\url{http://sirrah.troja.mff.cuni.cz/~mira/xitau/}}
\citep{Nemravova_2016A.A...594A..55N,Broz_2017ApJS..230...19B,Broz_2021A&A...653A..56B,Marchis_2021A&A...653A..57M,Broz_2022A&A...666A..24B,Broz_2022A&A...657A..76B,Ferrais_2022A&A...662A..71F,Broz_2023A&A...676A..60B,Fuksa_2023A&A...677A.189F,Oplistilova_2023A&A...672A..31O}.
It accounts for
mutual, N-body perturbations,
relativistic effects $\vec f_{\rm ppn}^i$ in the parametrized post-Newtonian (PPN) approximation
\citep{Standish_2006},
oblateness $\vec f_{\rm oblat}^i$ \citep{Kaula_1966tsga.book.....K},
as well as tides $\vec f_{\rm tide}^i$
\citep{Mignard_1979M&P....20..301M}.
The corresponding equation of motion
for the position vector $\vec r_i$ of the $i$-th component
in the barycentric reference frame
($\vec r_{ij} \equiv \vec r_i - \vec r_j$,
$m_j$ mass of the $j$-th component)
\begin{equation}
\ddot{\vec r}_i = -\sum_{j\ne i} {Gm_j\over r_{ij}^3}\vec r_{ij} + \vec f_{\rm ppn}^i + \vec f_{\rm oblat}^i + \vec f_{\rm tide}^i
\end{equation}
is integrated numerically, using the Bulirch-Stoer algorithm
\citep{Bulirsch_1966NuMat...8....1B,Levison_Duncan_1994Icar..108...18L},
with adaptive time stepping and the relative precision
$\varepsilon = 10^{-8}$.
Since the system has 4 components,
we have in total 48 dynamical and radiative parameters
(see Table~\ref{params} notes).
For minimisation of $\chi^2$, we use the simplex
\citep{Nelder_Mead_1965}
or subplex
\citep{Rowan_1990}
algorithms.

To constrain the model, one should use all kinds of observations
by minimising the metric
\begin{eqnarray}
\chi^2 &=& \chi^2_{\rm sky}
+ \chi^2_{\rm rv}
+ \chi^2_{\rm etv}
+ \chi^2_{\rm ecl}
+ \chi^2_{\rm vis}
+ \chi^2_{\rm clo}
+ \chi^2_{\rm t3}
+ \chi^2_{\rm lc} + \nonumber\\
&+& \chi^2_{\rm syn}
+ \chi^2_{\rm sed}
+ \chi^2_{\rm sed2}\,,
\end{eqnarray}
where individual terms are standard $O-C$'s squared,
divided by uncertainties~$\sigma$ squared, for
astrometry (SKY),
radial velocities (RV),
minima timings (ETV),
eclipse durations (ECL),
visibility (VIS),
closure phase (CLO),
triple product (T3),
light curves (LC),
normalized spectra (SYN),
spectral-energy distribution (SED),
and also relative brightness of the 4th component (SED2).

It is, however, often impossible to converge all parameters at once.
We thus proceeded sequentially,
by subsequently minimising the metrics
\begin{equation}
\chi^2 = \chi^2_{\rm etv} + \chi^2_{\rm ecl}\,,\label{eq2}
\end{equation}
\begin{equation}
\chi^2 = \chi^2_{\rm sky} + \chi^2_{\rm rv} + \chi^2_{\rm etv} + \chi^2_{\rm ecl} + \chi^2_{\rm sed} + \chi^2_{\rm sed2}\,,
\end{equation}
\begin{equation}
\chi^2 = \chi^2_{\rm sky} + \chi^2_{\rm rv} + \chi^2_{\rm etv} + \chi^2_{\rm ecl} + \chi^2_{\rm vis} + \chi^2_{\rm clo} + \chi^2_{\rm t3} + \chi^2_{\rm sed} + \chi^2_{\rm sed2}\,,
\end{equation}
always carefully considering which parameters should be free vs. fixed.
For instance, when using Eq.~(\ref{eq2}),
all radiative parameters must be fixed,
because they would be totally unconstrained.
Also, when some parameters are correlated,
like the total mass~$m_{\rm sum}$ and the distance~$d$
\citep{Kepler_1619ikhm.book.....K},
it is very useful to optimize grids of models,
in order to find the global, best-fit solution
and {\em eodem tempore\/} exclude other solutions.



\subsection{Reference model ((Aa+Ab)+B)+C} \label{3.1}

The reference model was adopted from
\citet{Nemravova_2016A.A...594A..55N}.
We added new observations though,
in a simplified form
---
RVs instead of full spectra,
minima timings instead of full light curves.
This allowed us to first focus on the dynamical (not radiative) parameters.
We also used a slightly different set of parameters
($P$ instead of $a$,
$\log e$ instead of $e$,
$\varpi$ instead of $\omega$,
$\lambda$ instead of $M$).

\begin{figure}
\centering
\includegraphics[width=8.3cm]{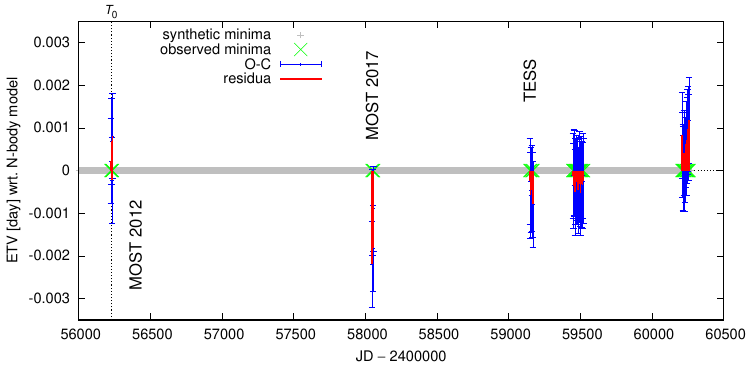}
\includegraphics[width=8.3cm]{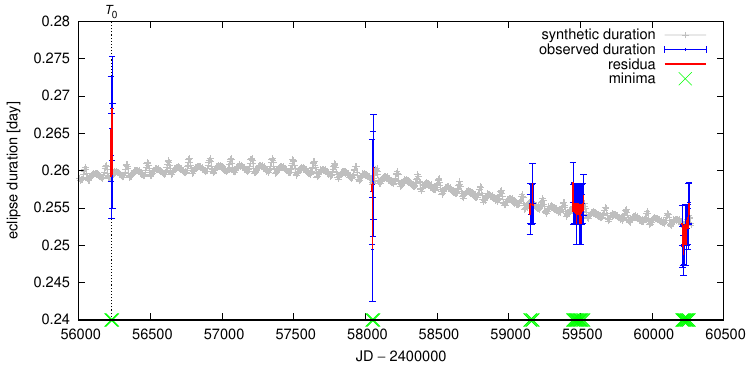}
\includegraphics[width=8.3cm]{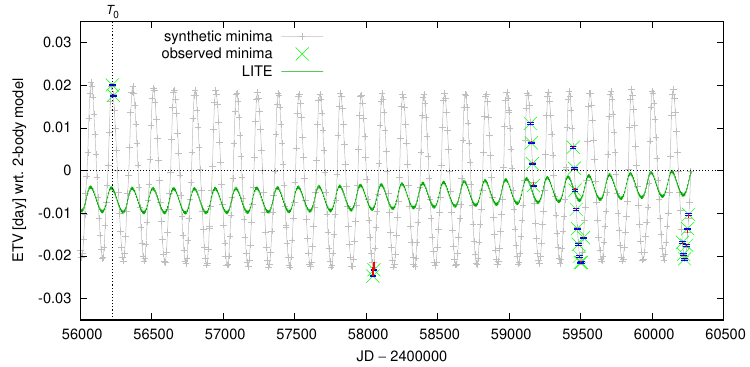}
\caption{
The reference model of $\xi$~Tau,
showing ETVs (top),
eclipse durations (middle),
perturbations with respect to a 2-body ephemeris (bottom),
MOST and TESS observations (\color{blue}blue\color{black}),
model (\color{gray}gray\color{black}),
residuals (\color{red}red\color{black}),
and for reference, also observed minima (\color{green}green\color{black}).
The total $\chi^2 = 1553$
(see Table~\ref{params} for individual contributions).
The mean ephemeris used in the bottom panel is
$P = 7.146665\,{\rm d}$, $T_0 = 2456224.704705$.
Our model is certainly able to explain the major perturbations,
which are substantial (${\pm}0.02\,{\rm d}$, or ${\pm}30\,{\rm min}$),
but some systematics remain to be explained
(up to 0.002\,d, or 3\,min).
}
\label{reference}
\end{figure}

\paragraph{Mirror solutions.}
We examined 8 mirror solutions with ETVs,
which are very sensitive to mutual perturbations.
We also used eclipse durations (ECL),
to constrain the precession.
Specifically, for the inclinations $i_1$, $i_2$,
and the longitudes of nodes $\Omega_1$, $\Omega_2$,
we tested the following combinations of discrete values
$86.5, 87.0, 93.0, 93.5^\circ$ and
$147.5, 148.5, 327.5, 328.5^\circ$, respectively.
Other orbital parameters were free,
except for the total mass $m_{\rm sum}$.
There is only one combination, which explains the ETVs
(Table~\ref{params}).
Their overall amplitude with respect to a two-body model
is about 0.02\,d
\citep{Nemravova_2016A.A...594A..55N},
although the residuals with respect to the N-body model
are ${<}0.002\,{\rm d}$
(Fig.~\ref{reference}).
The observed eclipse duration is decreasing
from 0.27 to 0.25\,d, and
the observed eclipse depth
is also decreasing from 0.110 to 0.077\,mag.
\vskip\baselineskip


However, ETVs are not enough to obtain all dynamical parameters.
We thus considered six datasets (SKY, RV, ETV, ECL, SED, SED2)
re-converged the reference model,
and obtained the free parameter values in
Table~\ref{params}
and the dependent parameter values in
Table~\ref{dependent}.
The unreduced $\chi^2_{\rm etv}$, $\chi^2_{\rm ecl}$ values
are already comparable to the number of degrees of freedom,
$\nu = N-M$,
on the other hand,
$\chi^2_{\rm rv}$ is relatively larger,
due to remaining systematics in RVs,
of the order of less than $1\,{\rm km}\,{\rm s}^{-1}$.
This, however, might be related to systematics of RVs,
which will be circumvented by using full spectra.

\subsection{Models constrained by interferometry} \label{3.2}


Next, we computed a grid of models
including three more, interferometric datasets
(VIS, CLO, T3).
Our aim was to constrain the total mass $m_{\rm sum}$ and the distance~$d$
and to exclude other, alternative solution.
We used 121 pairs of $m_{\rm sum}$ and $d$ values,
fixed these parameters,
but all other parameters were set free,
each model was re-converged,
so that it adapts 'at all cost'.

\begin{figure*}[p]
\centering
\includegraphics[width=16cm]{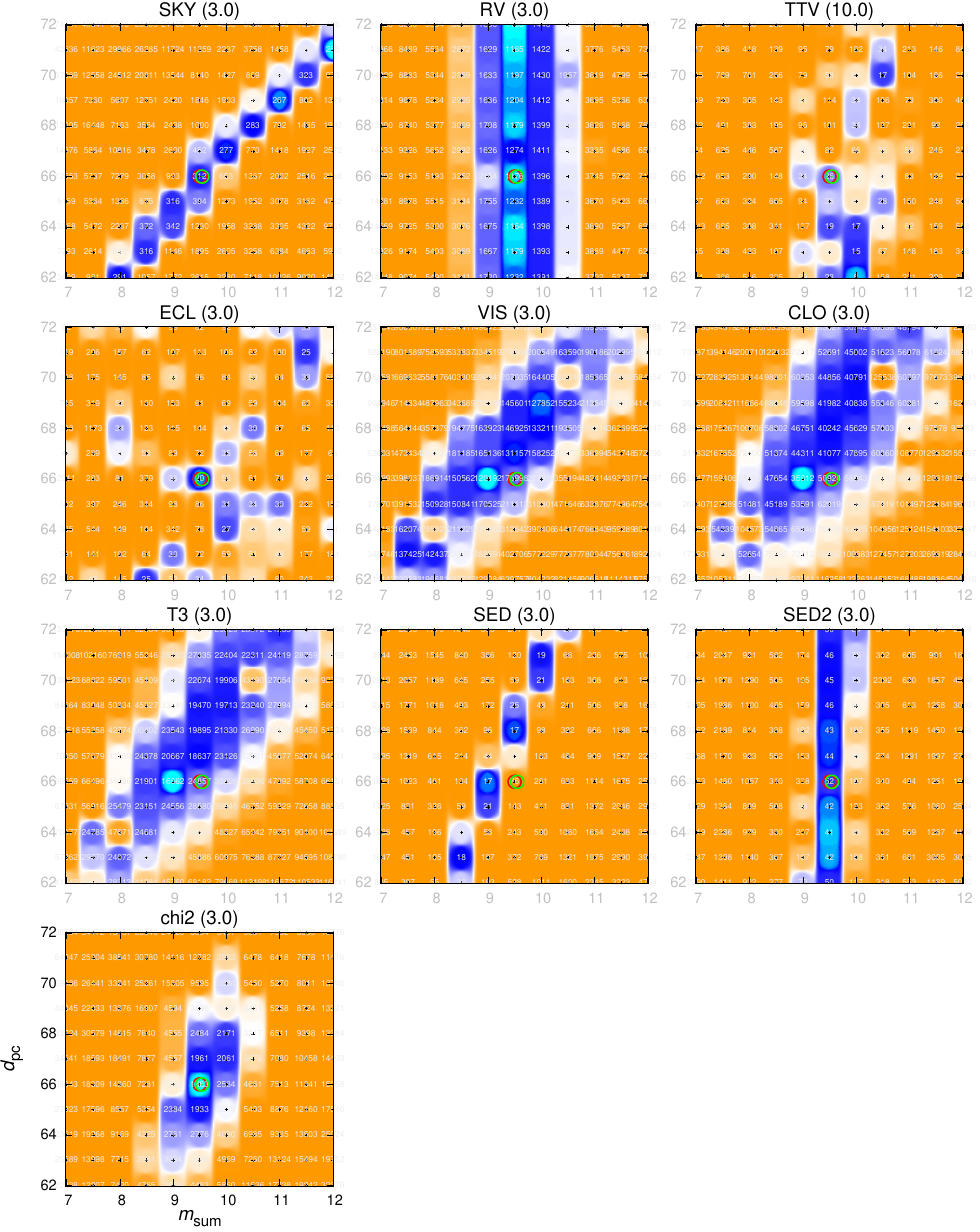}
\caption{
Best-fit models of $\xi$~Tau constrained by multi-technique observations
for two fixed parameters,
the total mass $m_{\rm sum}$ and the distance $d$.
The $\chi^2$ values (colours) indicate
best fit (\color{cyan}cyan\color{black}),
good fits (\color{blue}blue\color{black}),
poor fits (\color{orange}orange\color{black});
overall best fit (\color{red}red\color{black}) has $\chi^2 = 1661$ (unreduced).
We tested $121$ pairs of parameters,
$3000$ sequential iterations of simplex,
$3.6\times 10^5$ models in total,
other dynamical parameters (except $m_{\rm sum}$, $d$) were free
other radiative parameters were fixed
\citep{Nemravova_2016A.A...594A..55N}.
We constrained the model by
SKY, RV, ETV, ECL, SED, SED2 datasets; we didn't use
interferometry (but we used SKY),
spectroscopy (but RV), or
light curves (but ETV, ECL).
We assumed unit weights,
except $w_{\rm etv} = w_{\rm ecl} = 100$,
because we wanted to exclude all models with incorrect dynamics.
The two parameters are often positively correlated
(e.g. SKY, VIS, CLO, T3),
as predicted \citep{Kepler_1619ikhm.book.....K},
but some datasets are distance-independent or orthogonal
(RV, ETV, SED, SED2),
and the model is considered well-constrained.
after re-convergence (\color{green}green\color{black})
$\chi^2$ further decreased to $1553$.
The Gaia distance 59.6\,pc is excluded.
Additionally, VIS, CLO, T3 were also computed,
but their weights were set to zero,
otherwise our model would exhibit tension between VIS and CLO.
}
\label{msum_dpc_ALL}
\end{figure*}

\begin{figure*}[p]
\centering
\begin{tabular}{cc}
SKY & RV \\
\includegraphics[width=4cm]{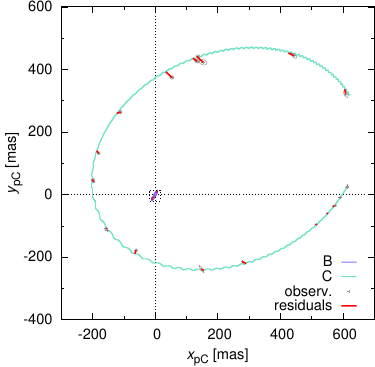}
\includegraphics[width=4cm]{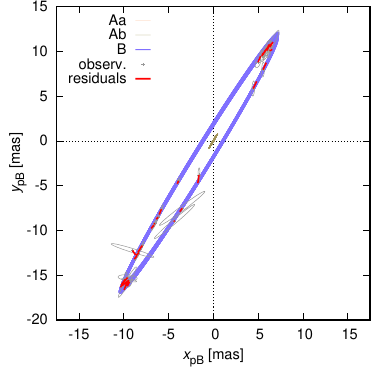} &
\includegraphics[width=8cm]{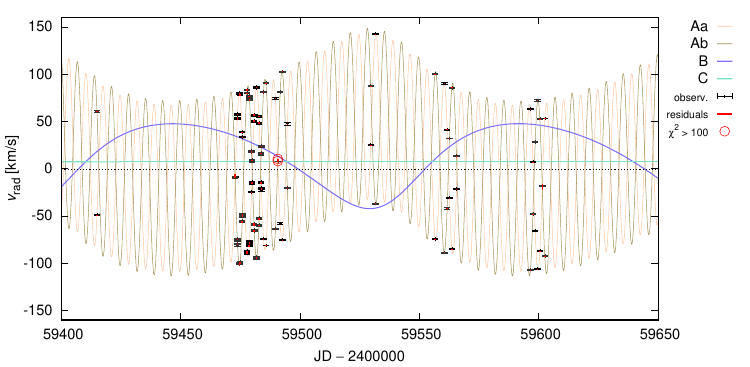} \\
ETV & ECL \\
\includegraphics[width=8cm]{figs/fitting17_NOVIS_SINGLE__1553/chi2_TTV.pdf} &
\includegraphics[width=8cm]{figs/fitting17_NOVIS_SINGLE__1553/chi2_ECL.pdf} \\
VIS & CLO, T3 \\
\includegraphics[width=8cm]{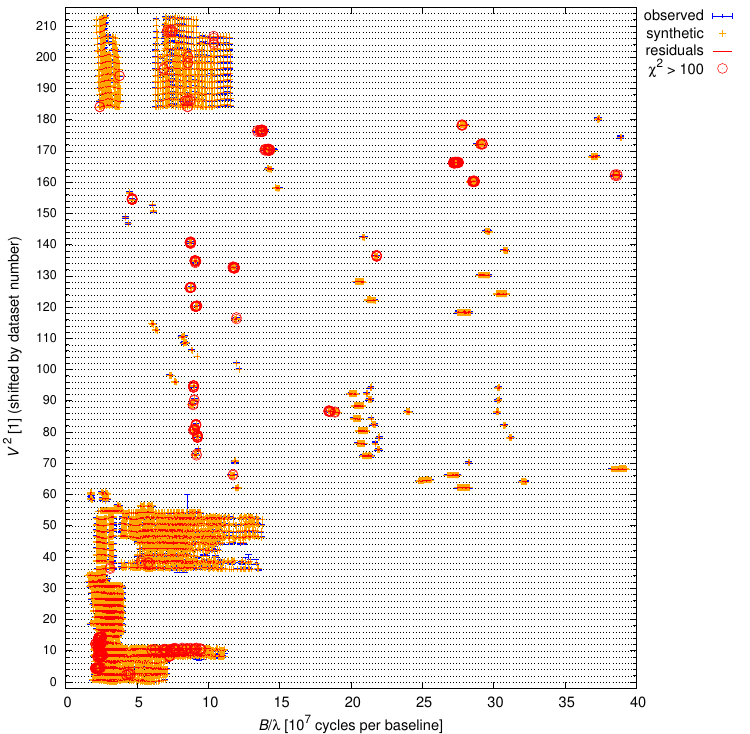} &
\hbox{\vbox{\hsize=8cm\hbox{%
\includegraphics[width=8cm]{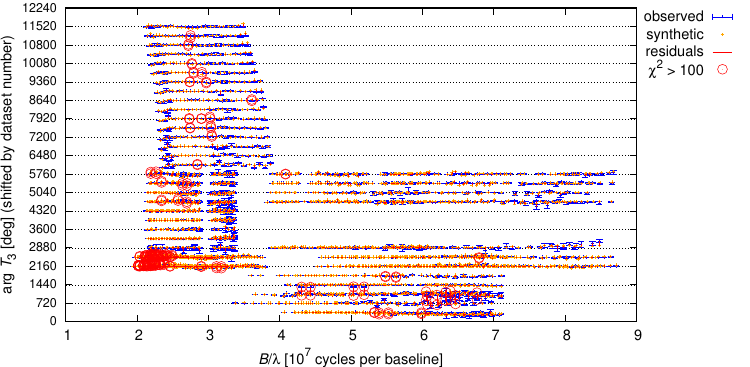}%
}\hbox{%
\includegraphics[width=8cm]{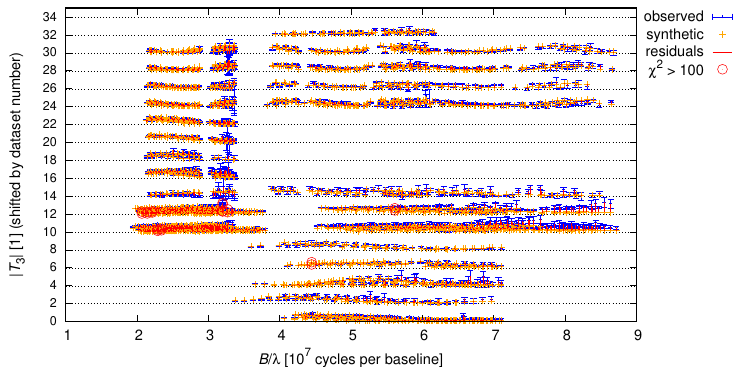}%
}}}
\\
SED & SED2 \\
\includegraphics[width=8cm]{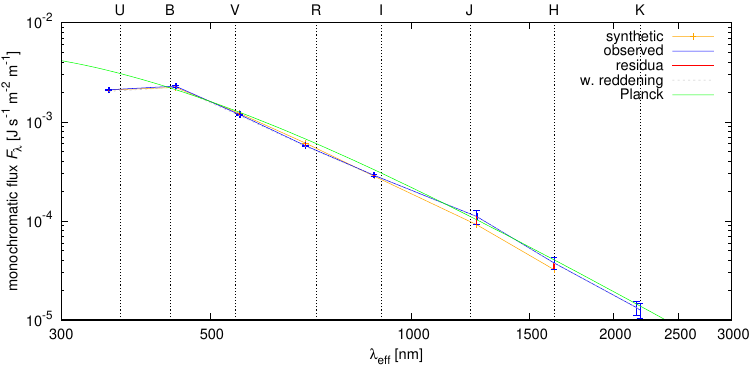} &
\includegraphics[width=8cm]{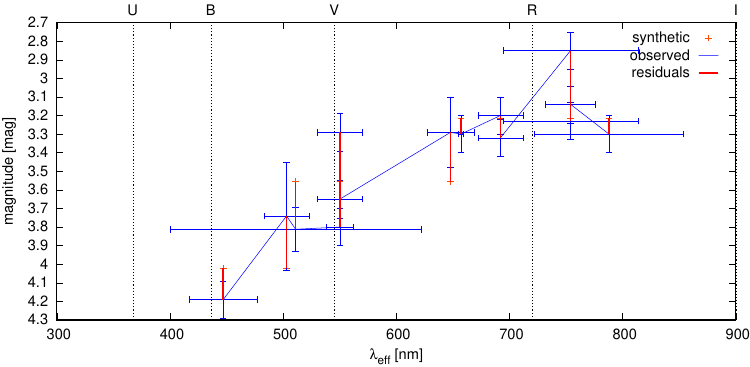} \\
\end{tabular}
\caption{
The best-fit model from Fig.~\ref{msum_dpc_ALL}
with $\chi^2 = 1553$ (unreduced).
Plots for each dataset,
SKY, RV, ETV, ECL, VIS, CLO, T3, SED, and SED2,
contain
observations (\color{blue}blue\color{black}),
synthetic data (\color{orange}yellow\color{black}), and
residuals (\color{red}red\color{black}).
The best-fit distance is
$d = 66.0\,{\rm pc}$.
One can notice some systematics for
ETV (cf. MOST 2017),
VIS, or
CLO datasets.
}
\label{fitting17_NOVIS_SINGLE__1553}
\end{figure*}

The results in Fig.~\ref{msum_dpc_ALL} show
logical correlations between $m_{\rm sum}$, $d$,
as required by astrometric or interferometric datasets
(i.e. angular separations of the 3rd and 4th components).
Yet other datasets
(RV, ETV, SED, SED2)
are orthogonal to this
by setting some scales absolutely,
in ${\rm km}\,{\rm s}^{-1}$, days, ${\rm W}\,{\rm m}^{-2}\,{\rm m}^{-1}$,
respectively.
This implies a uniqueness of the best-fit solution
(Table~\ref{params}).
How this solution with $\chi^2 = 1553$ looks like graphically is shown in 
Fig.~\ref{fitting17_NOVIS_SINGLE__1553}.
In either case, $\xi$~Tau distance must be around 66.0\,{\rm pc};
the Gaia distance 59.6\,pc is excluded.


\begin{table*}[p]
\caption{
Best-fit parameters of $\xi$~Tau models.
The `original' one is taken from \cite{Nemravova_2016A.A...594A..55N}, Table~15.
}
\centering
\tiny
\begin{tabular}{lllllllll}
                &                          & original  & reference & {\bf all-data} & tides      & 5-component &          \\
var.            & unit                     & val.      & val.      & val.           & val.       & val.        & $\sigma$ \\
\hline
\vrule height 8pt width 0pt
$m_{\rm sum}$   & $M_\odot$                & 8.881594  & 9.547810    & 9.549357    & 9.557040    & 10.410026   & 0.2      \\ 
$q_1$           & 1                        & 0.900146  & 0.948223    & 0.948717    & 0.948208    & 0.948885    & 0.001    \\ 
$q_2$           & 1                        & 0.880155  & 0.852451    & 0.854879    & 0.852786    & 0.876618    & 0.01     \\ 
$q_3$           & 1                        & 0.113367  & 0.175785    & 0.163744    & 0.177839    & 0.174983    & 0.01     \\ 
$q_4$           & 1                        &           &             &             &             & 0.080957    & 0.01     \\
$P_1$           & d                        & 7.148234  & 7.147145    & 7.147156    & 7.147136    & 7.147000    & 0.000010 \\ 
$\log e_1$      & 1                        & $-\infty$ & $-2.8400$   & $-2.8525$   & $-2.8627$   & $-2.8212$   & 0.01     \\ 
$i_1$           & $^\circ$                 & 86.530    & 86.702      & 87.185      & 86.847      & 87.120      & 0.1      \\ 
$\Omega_1$      & $^\circ$                 & 331.364   & 328.461     & 328.368     & 328.451     & 328.428     & 1.0      \\ 
$\varpi_1$      & $^\circ$                 & 245.479   & 113.967     & 116.317     & 119.557     & 120.890     & 1.0      \\ 
$\lambda_1$     & $^\circ$                 & 61.501    & 58.399      & 58.359      & 58.369      & 58.208      & 0.1      \\ 
$P_2$           & d                        & 145.744   & 145.792     & 145.785     & 145.784     & 145.772     & 0.01     \\ 
$\log e_2$      & 1                        & $-0.7044$ & $-0.6829$   & $-0.6829$   & $-0.6829$   & $-0.6826$   & 0.01     \\ 
$i_2$           & $^\circ$                 & 86.676    & 86.546      & 86.554      & 86.598      & 86.655      & 1.0      \\ 
$\Omega_2$      & $^\circ$                 & 328.902   & 328.347     & 328.349     & 328.347     & 328.444     & 1.0      \\ 
$\varpi_2$      & $^\circ$                 & 338.521   & 341.105     & 341.108     & 341.106     & 341.447     & 1.0      \\ 
$\lambda_2$     & $^\circ$                 & 64.198    & 63.587      & 63.587      & 63.587      & 63.811      & 1.0      \\ 
$P_3$           & d                        & 18503.8   & 18876.8     & 18877.1     & 18878.9     & 18887.0     & 100      \\ 
$\log e_3$      & 1                        & $-0.2451$ & $-0.2421$   & $-0.2396$   & $-0.2420$   & $-0.2417$   & 0.01     \\ 
$i_3$           & $^\circ$                 & $-26.275$ & $-23.748$   & $-23.031$   & $-23.820$   & $-23.956$   & 1.0      \\ 
$\Omega_3$      & $^\circ$                 & 108.252   & 101.844     & 102.066     & 102.121     & 101.376     & 1.0      \\ 
$\varpi_3$      & $^\circ$                 & 117.247   & 117.918     & 117.949     & 117.901     & 117.994     & 1.0      \\ 
$\lambda_3$     & $^\circ$                 & 148.550   & 148.445     & 148.334     & 148.433     & 148.450     & 1.0      \\ 
$T_1$           & K                        &           & 10540       & 11000       & 10540       & 10555       & 100      \\ 
$T_2$           & K                        &           & 10318       & 10602       & 10318       & 10318       & 100      \\ 
$T_3$           & K                        &           & 14046       & 14000       & 14046       & 13983       & 100      \\ 
$T_4$           & K                        &           & 6751$\dag$  & 6487$\dag$  & 6800$\dag$  & 6737$\dag$  & 300      \\ 
$T_5$           & K                        &           &             &             &             & 6737$\dag$  & 300      \\ 
$\log g_1$      & 1                        &           & 4.364       & 4.337       & 4.364       & 4.355       & 0.01     \\ 
$\log g_2$      & 1                        &           & 4.343       & 4.355       & 4.343       & 4.343       & 0.01     \\ 
$\log g_3$      & 1                        &           & 4.300       & 4.238       & 4.300       & 4.257       & 0.01     \\ 
$\log g_4$      & 1                        &           & 4.240$\dag$ & 4.258$\dag$ & 4.237$\dag$ & 4.241$\dag$ & 0.01     \\ 
$\log g_5$      & 1                        &           &             &             &             & 4.241$\dag$ & 0.01     \\ 
$v_{\rm rot,1}$ & ${\rm km}\,{\rm s}^{-1}$ &           &             & 14.600      & 15.203      & 15.203      & 10       \\ 
$v_{\rm rot,2}$ & ${\rm km}\,{\rm s}^{-1}$ &           &             & 13.281      & 15.210      & 15.210      & 10       \\ 
$v_{\rm rot,3}$ & ${\rm km}\,{\rm s}^{-1}$ &           &             & 242.262     & 242.371     & 242.371     & 10       \\ 
$v_{\rm rot,4}$ & ${\rm km}\,{\rm s}^{-1}$ &           &             & 65.234      & 65.234      & 65.234      & 30       \\ 
$v_{\rm rot,5}$ & ${\rm km}\,{\rm s}^{-1}$ &           &             &             &             & 65.234      & 30       \\ 
$\Delta t_1$    & s                        &           &             &             & 800         & 100         & 100      \\ 
$\Delta t_2$    & s                        &           &             &             & 800         & 100         & 100      \\ 
$C_{20,1}$      & 1                        &           &             &             & 0.000       & 0.000       & 0.001    \\ 
$C_{20,2}$      & 1                        &           &             &             & 0.000       & 0.000       & 0.001    \\ 
$\gamma$        & ${\rm km}\,{\rm s}^{-1}$ & 8.82      & 10.180      & 11.022      & 9.870       & 9.625       & 0.5      \\ 
$d$             & pc                       & 67.88     & 65.994      & 66.138      & 66.015      & 67.947      & 1.0      \\
\hline
\vrule height 8pt width 0pt
$n_{\rm sky}$       & & 96   & 106   & 106     & 106   & 106   \\
$n_{\rm rv}$        & & 834  & 260   & 822     & 1103  & 1103  \\
$n_{\rm etv}$       & & 35   & 48    & 77      & 77    & 77    \\
$n_{\rm ecl}$       & & 4    & 47    & 47      & 47    & 47    \\
$n_{\rm vis}$       & & --   & 16678 & 16678   & 16678 & 16678 \\
$n_{\rm clo}$       & & --   & 4533  & 4533    & 4533  & 4533  \\
$n_{\rm t3}$        & & --   & 4117  & 4117    & 4117  & 4117  \\
$n_{\rm lc}$        & & --   & --    & 18489   & --    & --    \\
$n_{\rm syn}$       & & --   & --    & 1753731 & --    & --    \\
$n_{\rm sed}$       & & --   & 7     & 7       & 7     & 7     \\
$n_{\rm sed2}$      & & --   & 14    & 14      & 14    & 14    \\
$n$                 & & 978  & 482   & 1798621 & 1354  & 1354  \\
\hline                                                      
\vrule height 8pt width 0pt
$\chi^2_{\rm sky}$  & & 185  & 301    & 324     & 289         & 233         \\
$\chi^2_{\rm rv}$   & & 2237 & 1152   & 3686    & 3377        & 3286        \\
$\chi^2_{\rm etv}$  & & 151  & 20.2   & 76.6    & 392$^\ddag$ & 453$^\ddag$ \\
$\chi^2_{\rm ecl}$  & & 3.3  & 20.3   & 144!    & 21.1        & 43.4        \\
$\chi^2_{\rm vis}$  & & --   & 174346 & 160495  & 174207      & 120467      \\
$\chi^2_{\rm clo}$  & & --   & 62129  & 42703   & 61131       & 48651       \\
$\chi^2_{\rm t3}$   & & --   & 26618  & 23463   & 25681       & 18191       \\
$\chi^2_{\rm lc}$   & & --   & --     & 160870  & --          & --          \\
$\chi^2_{\rm syn}$  & & --   & --     & 3246892 & --          & --          \\
$\chi^2_{\rm sed}$  & & --   & 18.5   & 297!    & 18.8        & 28.8        \\
$\chi^2_{\rm sed2}$ & & --   & 40.9   & 224!    & 46.3        & 40.3        \\
$\chi^2$            & & 2578 & 1553   & 3639180 & 4145        & 4085        \\
\end{tabular}
\tablefoot{
The notation is as follows:
$m_i \hbox{ for }\forall i$ the masses,
$q_1 = m_2/m_1$,
$q_2 = m_3/(m_1+m_2)$,
$q_3 = m_4/(m_1+m_2+m_3)$ mass ratios,
$P_i$ period,
$\log e_i$ logarithm of eccentricity,
$i_i$ inclination,
$\Omega_i$ longitude of the ascending node,
$\varpi_i$ longitude of the pericentre,
$\lambda_i$ mean longitude in orbit,
$T_i$ effective temperature,
$\log g_i$ logarithm of gravitational acceleration,
$v_{{\rm rot},i}$ rotational velocity,
$C_{20,i}$ oblateness,
$\Delta t_i$ tidal time lag,
$n$ number of observations,
$\chi^2$ unreduced \textipa{[kaI skwe@]} value;
$\chi^2/n$ would be reduced one;
$\dag$ constrained by mass using \cite{Harmanec_1988BAICz..39..329H};
$\ddag$ with ETV uncertainties of the order of $10^{-4}\,{\rm d} \simeq 0.1\,{\rm min}$.
The epoch of osculating elements is 2456224.724705 (TDB).
The angular elements are expressed in the observer's reference frame.
}
\label{params}
\end{table*}

\paragraph{MCMC analysis.}
For this model, we estimated the uncertainties of model parameters
with the Markov chain Monte Carlo (MCMC)
\citep{Markov_1906},
which should reflect the uncertainties of observational data.
Here, we used all dynamical parameters,
but not all radiative,
in order to prevent systematics from affecting the resulting uncertainties.
If any systematics are present
(e.g. tension between VIS and CLO)
it would result in very small, unrealistic uncertainties.
The standard corner plot is shown in
Fig.~\ref{mcmc17_NOVIS_SINGLE}.


\subsection{Models constrained by CHIRON and TESS observations} \label{3.3}


Next, we included
44 high-resolution spectra (from CHIRON, one per night) and
8 high-precision light curves (from MOST and TESS)
and computed another grid of models.
Our aim was to better constrain the radiative parameters,
in particular, the effective temperatures $T_1$ and $T_3$.
We again used 121 pairs of $T_1$ and $T_3$ values,
fixed them,
but all other parameters were free.

\begin{figure*}[htpb]
\centering
\includegraphics[width=16cm]{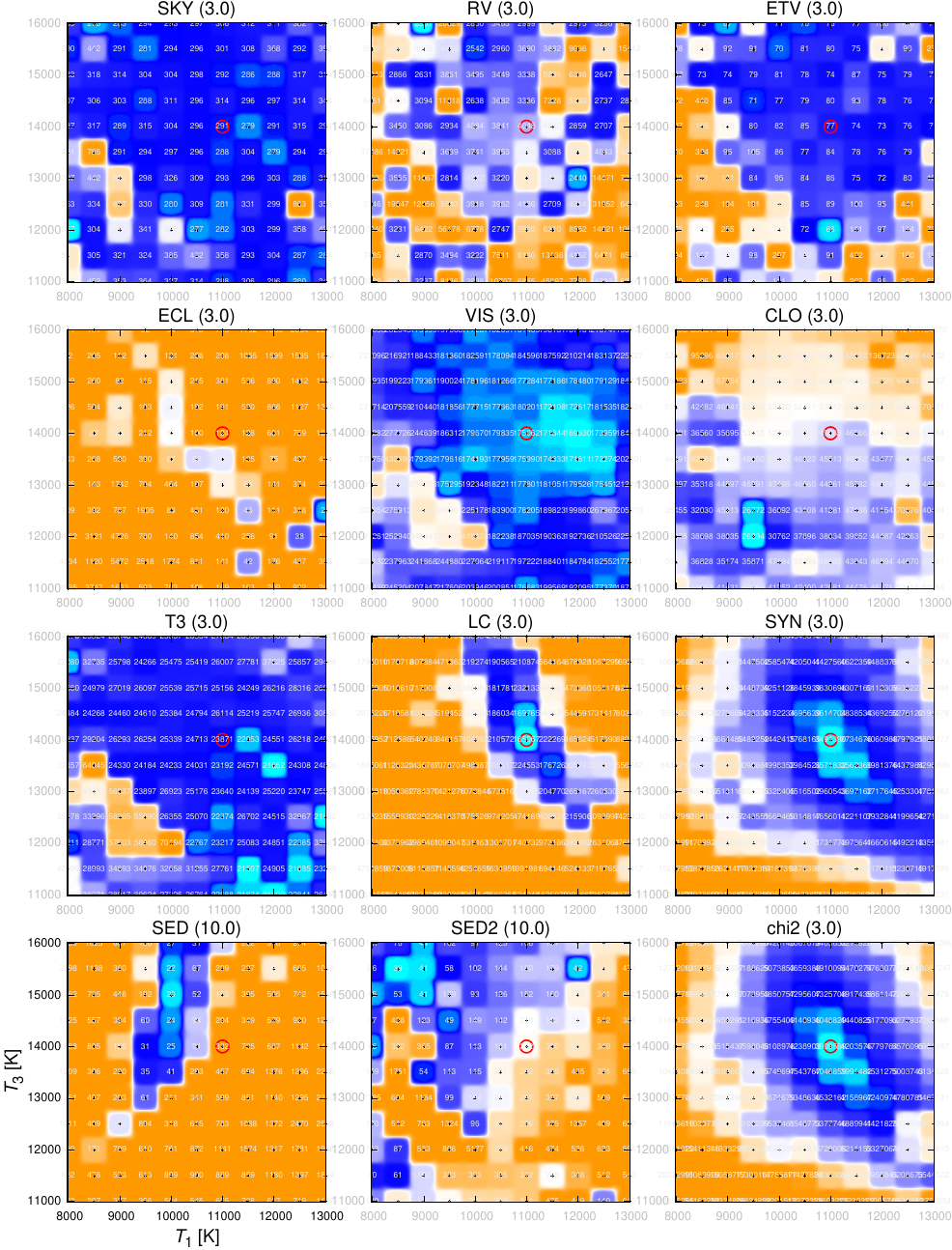}
\caption{
Same as Fig.~\ref{msum_dpc_ALL}
for two different fixed parameters,
effective temperatures $T_1$, $T_3$.
This time,
light curves (LC) and
synthetic spectra (SYN)
were included.
The overall best fit $\chi^2 = 3639180$ (unreduced). 
One can notice some systematics, e.g.
tension between VIS and CLO,
best ECL is offset,
best SED is offset, and
best SED2 is very offset.
}
\label{T1_T3_ALL}
\end{figure*}

The results in Fig.~\ref{T1_T3_ALL} show some datasets,
which are insensitive to the radiative parameters
(e.g. SKY, RV, ETV, ECL),
fine,
but others constrain the temperatures very well
(VIS, LC, SYN).
In particular, light curves and eclipse depths are very `strict' constraints
($T_1 \simeq 11000\,{\rm K}$, $T_3 \simeq 14000\,{\rm K}$),
fully in agreement with spectral line profiles.
The 3rd component does not eclipse anything,
but its brightness contributes as a 3rd light.
The line profiles of Aa, Ab, B components are blended,
but easy to disentangle
because of different rotation speeds
(Fig.~\ref{synthetic2_4861}).


\paragraph{Tension.}
However, looking at individual contributions to $\chi^2$
versus the position of the best-fit model
(Fig.~\ref{T1_T3_ALL}, red circle),
one can notice the following problems.
On one hand, CLO dataset exhibits tension
(with respect to VIS),
where better fits of CLO data require lower $T_3$.
On the other hand, SED, SED2 datasets exhibit different tension
(w.r.t. LC, SYN),
requiring higher $T_3$.
It is clearly not easy to correct these issues by changing $T_3$,
or, as a matter of fact, by changing any other parameter;
we note all other parameters were free.


\paragraph{Systematics.}
There might be various reasons for such inconsistencies.
First, our model could be internally inconsistent,
for example,
the light curve algorithm \citep{Wilson_1971ApJ...166..605W,Wilson_2010ApJ...723.1469W},
which is technically comparable to \citet{Prsa_2016ApJS..227...29P},
could be incompatible with grids of synthetic spectra
\citep{Lanz_2003ApJS..146..417L,Lanz_2007ApJS..169...83L,Palacios_2010A&A...516A..13P,Laverny_2012A&A...544A.126D,Husser_2013A&A...553A...6H},
used for fitting of the SYN dataset.
Second, some approximations might be inadequate,
for example, the 3rd component is a fast-rotating star,
but we used a synthetic spectrum suitable for slow-rotating stars.
The same is true for the interferometric quantities,
or VIS, CLO, T3,
where we assumed limb-darkened disks
\citep{Hanbury_1974MNRAS.167..475H}.
Third, some fixed parameters might be set incorrectly,
for example, the linear limb darkening coefficients,
for which we assumed standard values
\citep{vanHamme_1993AJ....106.2096V}.
Fourth, discretisation errors might be larger than expected,
especially for the ETV, ECL datasets,
where we used a linear interpolation between the neighboring time steps.
We checked for all of the possibilities above,
but none of them explains the tension.


\paragraph{Best fit.}
How the best-fit solution with $\chi^2 = 3639180$ looks like is shown in
Fig.~\ref{fitting19_teffs_SINGLE__3639180}.
The fit of the SKY dataset is acceptable;
new WDS data of the 4th component are fitted perfectly,
the NPOI data of the 3rd component exhibit only minor systematics.
Note the coordinates are photometric,
with respect to Aa+Ab, or Aa+Ab+B, respectively.
The RVs are also acceptable;
new CHIRON data are fitted perfectly,
in agreement with line profiles (SYN),
old RV data exhibit offsets up to 5\,km/s,
but not systematically.
For the ETVs, the offsets are ${<}0.001\,{\rm d}$,
but one can notice a trend that
MOST 2012 is fitted,
TESS is fitted,
but MOST 2017 is offset by $-0.0015\,{\rm d}$.
It is clearly not easy to correct these issues,
within the current dynamical model (see Sect.~\ref{tides}).
We verified that the transversal light-time effect
\citep{Conroy_2018ApJ...854..163C,Vokrouhlicky_2026}
has a substantially smaller amplitude.
Similar, related offsets are seen in the ECL dataset,
in agreement with light curves. 
The fits of VIS, CLO, T3 datasets contain individual
offsets for certain baselines $B/\lambda$,
but the trends of $V^2(B/\lambda)$ are correct.
This is often caused by visibility calibration,
or transfer function issues.
The synthetic SED dataset is overestimated (by 0.1 to 0.15\,mag),
while SED2 is underestimated (by ${\sim}0.3\,{\rm mag}$).
This might suggest an issue related to the 4th component,
the total mass $m_{\rm sum}$, or
the distance $d$,
but more likely to radiative parameters
(cf. the discussion of $T_3$).
The light curve fit seems to be almost perfect,
for both the primary and secondary minima,
but there are tiny, statistically significant offsets in minima timings,
corresponding 1:1 to the ETV offsets above.
The fit of normalized spectra (SYN) is also acceptable,
including the H$\alpha$, H$\beta$ wings,
with a notable exception of the H$\alpha$, H$\beta$ core depths
for Aa, Ab components,
which are slightly underestimated.
This might suggest an issue related to the
grids of synthetic spectra
or relative luminosities of the components (Aa, Ab, B).


\subsection{Models with tides and oblateness} \label{tides}

\begin{figure*}[htpb]
\centering
\begin{tabular}{cc}
\includegraphics[width=4cm]{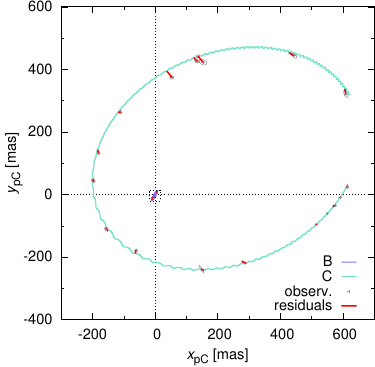}
\includegraphics[width=4cm]{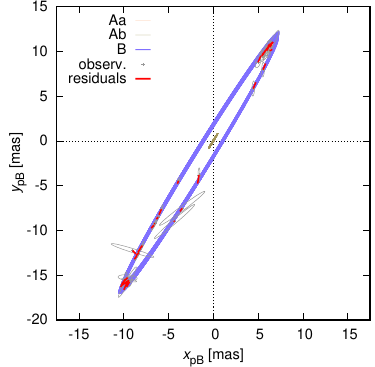} &
\includegraphics[width=8cm]{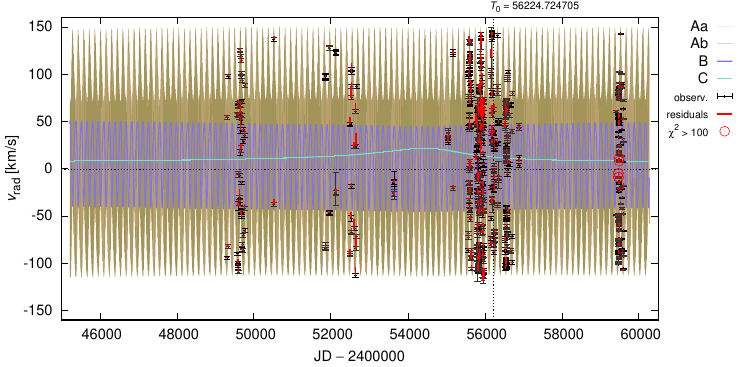} \\
\includegraphics[width=8cm]{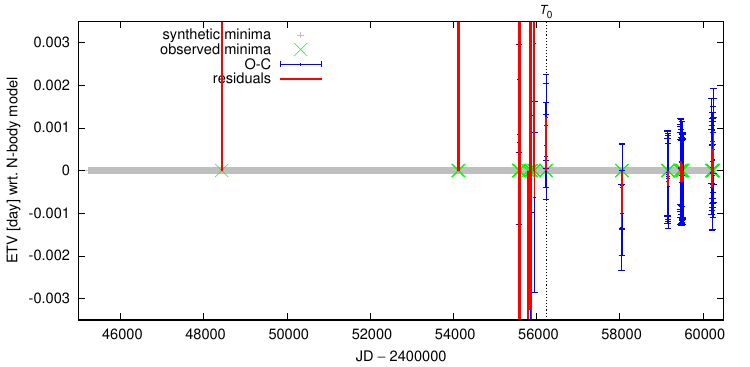} &
\includegraphics[width=8cm]{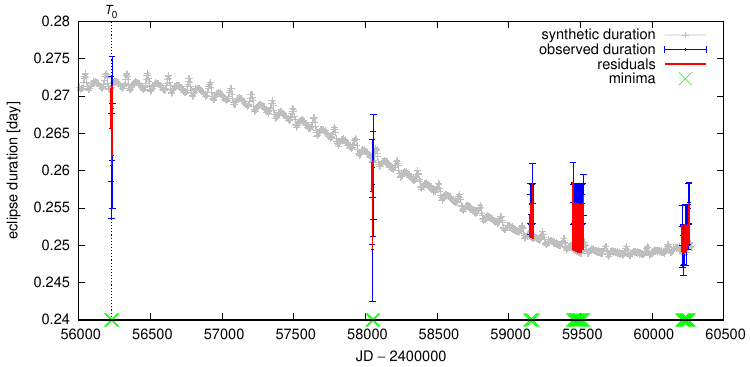} \\
\includegraphics[width=8cm]{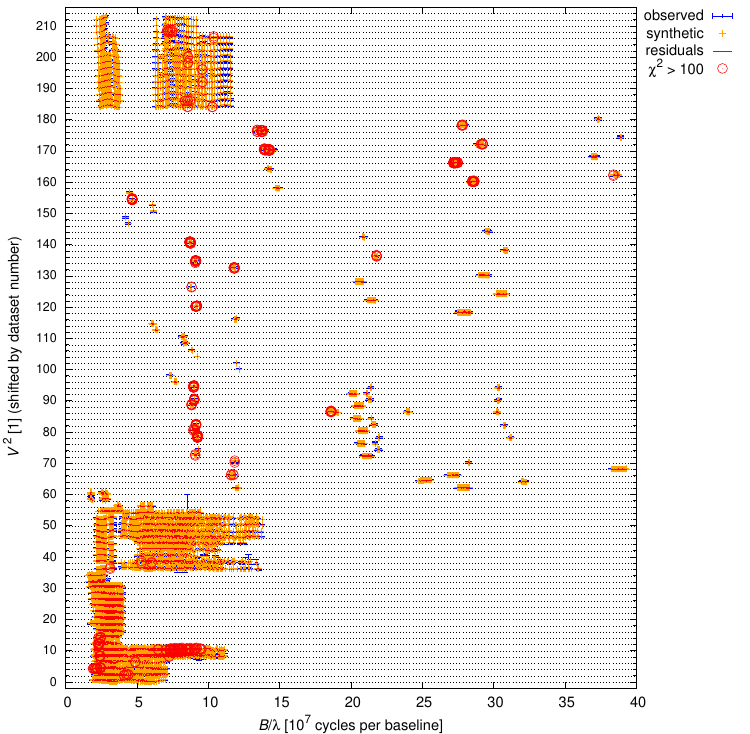} &
\hbox{%
\vbox{\hsize=8cm
\hbox{%
\includegraphics[width=8cm]{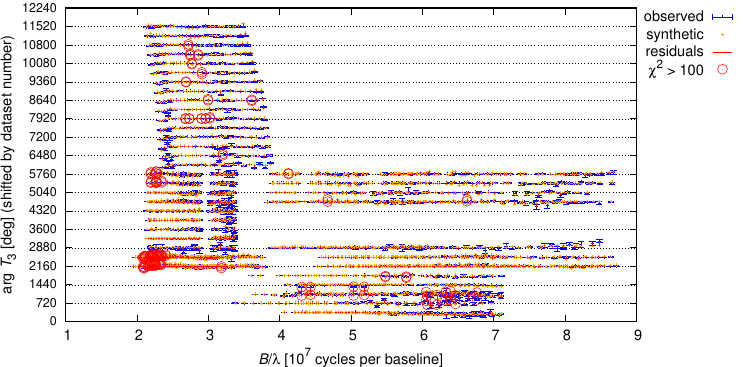}%
}\hbox{%
\includegraphics[width=8cm]{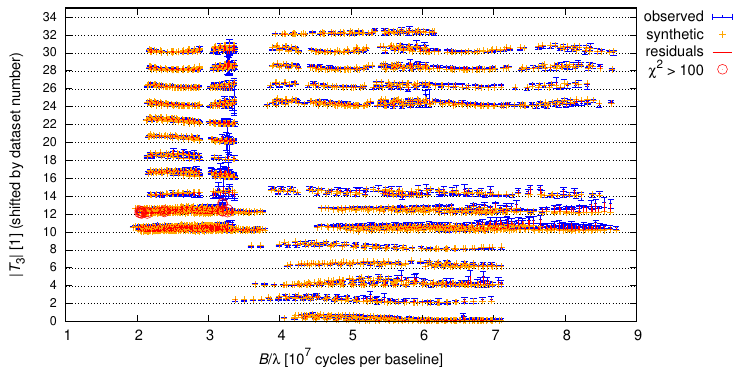}%
}%
}%
}%
\\
\includegraphics[width=8cm]{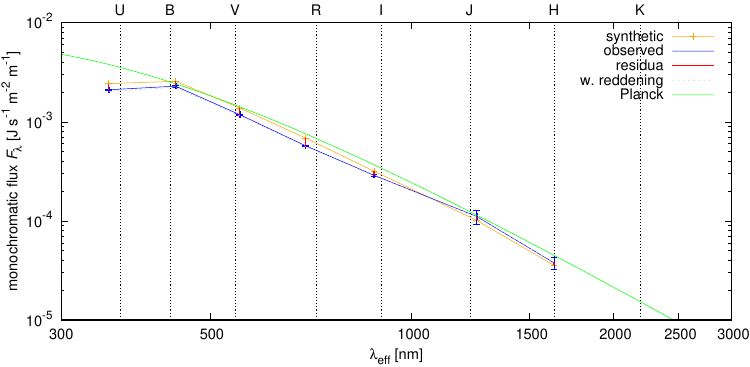} &
\includegraphics[width=8cm]{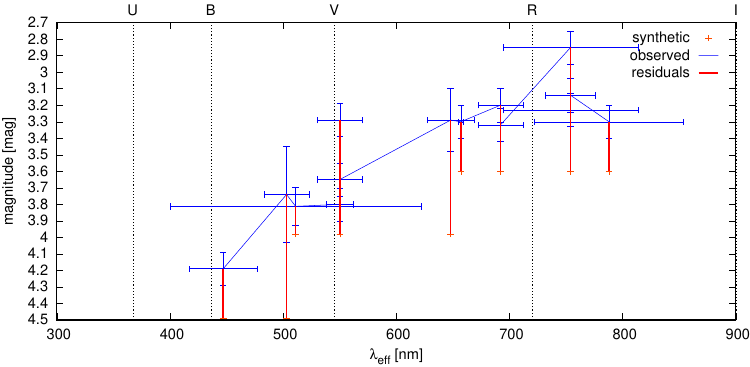} \\
\includegraphics[width=8cm]{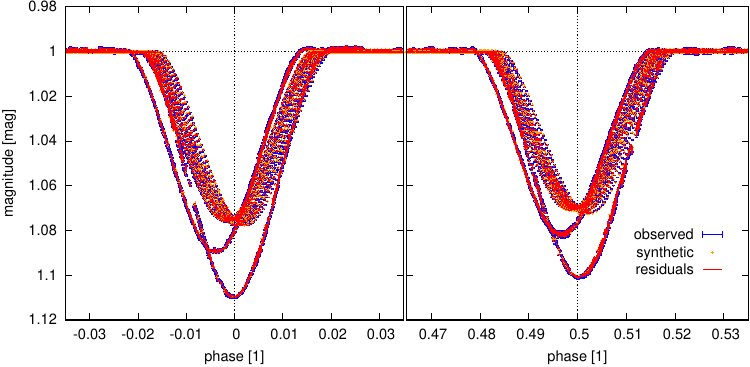} &
\includegraphics[width=8cm]{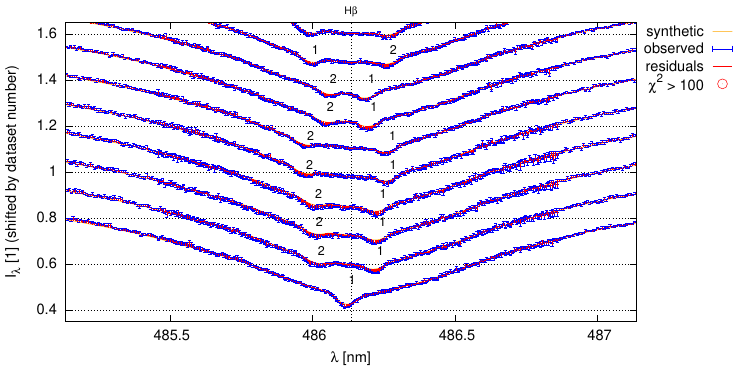} \\
\end{tabular}
\caption{
The best-fit model from Fig.~\ref{T1_T3_ALL}
with $\chi^2 = 3639180$ (unreduced).
The light curves (LC) were phased,
to show different minima depths (MOST vs. TESS).
The synthetic spectra (SYN) plot presents
only a subset of 10 spectra
and only the H$\beta$ line region.
One can still notice some systematics for
ETV (MOST 2017 offset),
ECL (TESS offset),
VIS (individual observations),
CLO ({\em ditto\/}),
SED (overestimated),
SED2 (underestimated), or
SYN (Aa, Ab line depths are not sufficient).
}
\label{fitting19_teffs_SINGLE__3639180}
\end{figure*}

In order to explain the trend seen in ETVs,
in particular, the negative offset $-0.0015\,{\rm d}$ of MOST 2017 timings
(Fig.~\ref{fitting19_teffs_SINGLE__3639180}),
we tested models with tides and oblateness.%
\footnote{In the context of stellar studies, these effects are commonly denoted equilibrium and non-equilibrium tides \citep[e.g.,][]{Hut_1981A&A....99..126H}.}
These two terms induce a change of the period ($\dot P$)
and an additional precession ($\dot\omega$, $\dot\Omega$),
added to the existing precession due to four bodies.
At the same time, we revised the uncertainties of ETVs to
$10^{-4}\,{\rm d}$,
in agreement with high-precision light curves
and our aim to explain offsets at this level.
We computed a grid of models parametrized by
the oblateness~$C_{20}$ and
the time lag~$\Delta t$
\citep{Mignard_1979M&P....20..301M}.
Only the closest components (Aa, Ab) are relevant,
as the remaining (B, C) are too distant
and the terms are negligible.
For simplicity, we also assumed no misalignment
and no spin precession.

\begin{figure*}
\centering
\includegraphics[width=12cm]{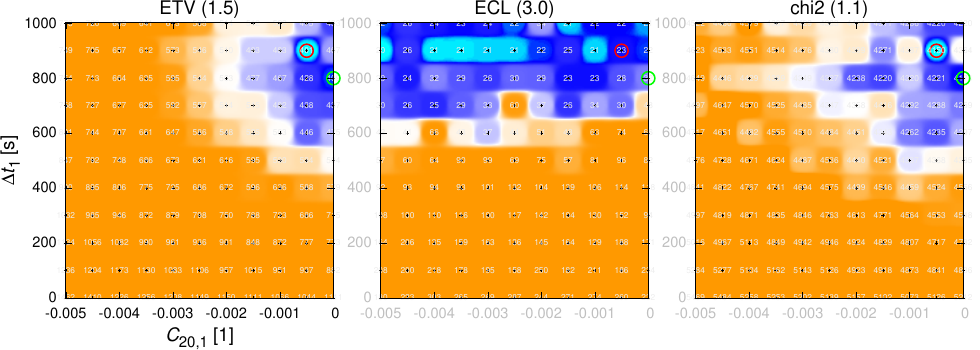}
\caption{
Same as Fig.~\ref{msum_dpc_ALL}
for different fixed parameters,
the oblateness $C_{20,1}$ and
the tidal time lag $\Delta t_1$.
For the 2nd component we assumed the same values.
The uncertainties of ETVs were revised to $10^{-4}\,{\rm d}$.
We tested 121 pairs of parameters,
300, 300, and 1000 sequential iterations of simplex,
$1.9\times 10^5$ models in total.
The parameters of the inner, Aa+Ab orbit (plus $P_2$) were free.
The best fit is indicated
(\color{red}red\color{black})
along with a statistically equivalent, close-to-zero $C_{20,1}$ solution
(\color{green}green\color{black}).
Zero $\Delta t_1$ is excluded.
}
\label{C201_Deltat1_TTV_3}
\end{figure*}

\begin{figure*}
\centering
\includegraphics[width=12cm]{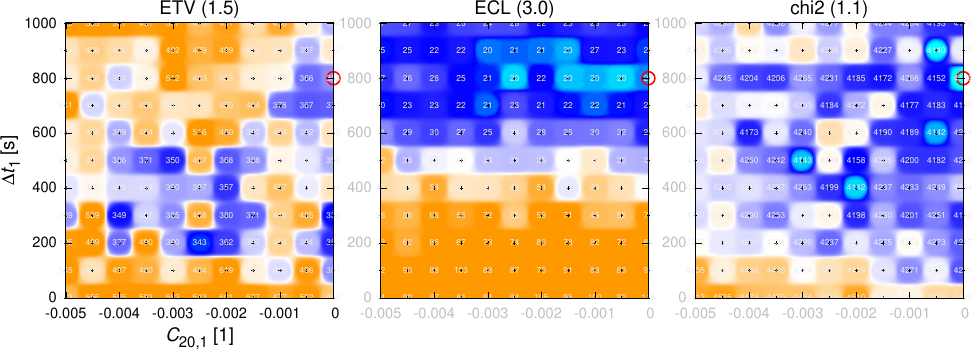}
\caption{
Same as Fig.~\ref{C201_Deltat1_TTV_3},
but the parameters of all orbits were free.
The solutions differ from Fig.~\ref{C201_Deltat1_TTV_3}
partly because of perturbations by component C.
The solutions with low $\Delta t_1 \sim 100\,{\rm s}$
exhibit a poor fit of eclipse durations (ECL),
so that the solution with high $\Delta t_1$ (and zero $C_{20,1}$)
is still preferred.
}
\label{C201_Deltat1_TTV_6}
\end{figure*}

The results in Fig.~\ref{C201_Deltat1_TTV_3} show
a clear signal for tides, but not for oblateness.
The time lag $\Delta t$ is about $(800\pm 100)\,{\rm s}$,
if only the 1st orbit was free.
If the 2nd and 3rd orbits were also free,
the time lag could be as low as ${\sim}100\,{\rm s}$,
but only at the expense of a poor fit of the ECL dataset
(Fig.~\ref{C201_Deltat1_TTV_6}).


\begin{figure*}
\centering
\includegraphics[width=12cm]{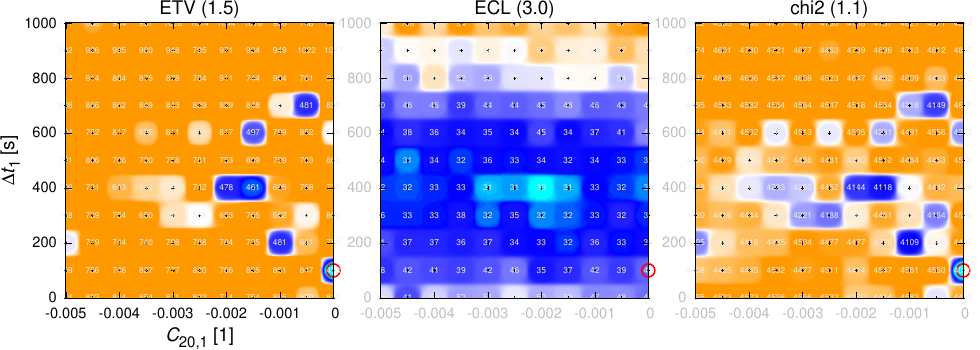}
\caption{
Same as Fig.~\ref{C201_Deltat1_TTV_3},
but for models with five components (Aa+Ab+B+Ca+Cb).
The perturbations and the light-time effect due to Ca+Cb components were stronger.
The best-fit solution with low $\Delta t_1 \sim 100\,{\rm s}$
now exhibits a good fit of eclipse durations (ECL).
}
\label{fitting22_allrvs_BINARY__4085}
\end{figure*}


Note we computed the tides for the reference
radius $R = R_\odot$,
the Love number $k_2 = 0.3$,
the mean motion $n \doteq 0.879\,{\rm rad}\,{\rm d}^{-1}$, and
the rotation frequency $\omega = 1.25\,{\rm rad}\,{\rm d}^{-1}$.
One can rescale $\Delta t$ to any radius, e.g., $R' = R_1$
(and $k_2' = 0.03$),
because the tidal term is proportional to $R^5 k_2\Delta t$, so
\begin{equation}
\Delta t' = \Delta t(R/R')^5(k_2/k_2') \simeq (710\pm 90)\,{\rm s}\,.
\end{equation}
One can convert it to the dissipation factor
\begin{equation}
Q = {1\over \Delta t\,2(\omega-n)} \simeq 160\,,
\end{equation}
or equivalently
\begin{equation}
Q' \equiv {Q\over k_2} \simeq 5500\,,
\end{equation}
or equivalently
\begin{equation}
E \equiv {2\over 3} k_2\Delta t\,2(\omega-n) \simeq 1.2\times 10^{-4}\,.
\end{equation}
This seems to be too high for non-rotating B-type stars
($\Delta t \simeq 0.15\,{\rm s}$,
$Q \simeq 10^6$,
$Q' \simeq 3\times 10^7$,
$E \simeq 2\times 10^{-8}$;
\citealt{Zahn_1975A&A....41..329Z,Zahn_1977A&A....57..383Z}).

From this point of view, the model with tides is excluded.
We point out just a few arguments,
why dissipation could be larger than expected.
Some stars orbited by exoplanets
have substantially smaller $Q' \simeq 5\times 10^4$
\citep{Penev_2018AJ....155..165P,Maciejweski_2016A&A...588L...6M}.
Rotating stars exhibit
Eddington--Sweet circulation
\citep{Eddington_1925Obs....48...73E,Sweet_1950MNRAS.110..548S},
Soldberg--H\o iland instability
\citep{Solberg_1936ApNr....1..237S,Hoiland_1941}, or
Goldreich--Schubert--Fricke instability
\citep{Goldreich_1967ApJ...150..571G,Fricke_1968ZA.....68..317F},
and such stars have smaller $Q'$.
Some binaries show signs of large-amplitude, resonant oscillation,
not only at the orbital frequency (`heartbeat'),
but at much higher multiples,
e.g. 229 times
\citep{Fuller_2017MNRAS.472L..25F,Fuller_2024MNRAS.527L.103F}.
Finally, the inner orbit (Aa+Ab) has non-zero, forced eccentricity,
$e \to 0.008$,
due to component B, which increases dissipation as on Io
($e = 0.004$; \citealt{Peale_1979Sci...203..892P}).


Alternative solutions also exist
for non-zero oblateness,
e.g., $C_{20,1} = C_{20,2} = -0.001$.
Whether or not it is reasonable
one can estimate from rotational and tidal (Roche) deformation.
Assuming the Love number
\citep{Love_1909RSPSA..82...73L}
of the order of
$k_2 \simeq 0.01$ to $0.1$
for radiative and convective stars, respectively
\citep{Claret_2004A&A...424..919C},
the oblateness from rotation is
\citep{Kaula_1966tsga.book.....K}
\begin{equation}
C_{20} = -{2\over 3} k_2 \left({\Omega\over n}\right)^2\,,
\end{equation}
and from tides
\begin{equation}
C_{20} = -{2\over 3} k_2 \left({R\over a}\right)^3\,.
\end{equation}
The maximal value (convective, rotational) is
${\sim}1.5\times 10^{-4}$.
From this point of view, the model with oblateness is also excluded.




\subsection{Models with five components ((Aa+Ab)+B)+(Ca+Cb)} \label{five}

Since some tension remains in the current model
--- specifically, high-precision astrometry
\citep{Tokovinin_2020AJ....160....7T},
leads to slightly more `elongated' outer orbit,
the RV dataset constrains the masses of components Aa, Ab, B,
the ETV dataset would require more massive component~C,
the SED dataset larger distance (69\,pc),
and the SED2 dataset a fainter component~C,
--- an elegant solution would be splitting of component~C
into a binary (Ca+Cb).
This decreases its luminosity
by up to ${\sim}2.25\,{\rm mag}$,
depending on the mass ratio.

We computed the same grid of models as in Sect.~\ref{tides}.
The results in Fig.~\ref{fitting22_allrvs_BINARY__4085} show 
a few solutions with non-zero tides,
$\Delta t = (100\pm 50)\,{\rm s}$,
which is more reasonable value than in Sect.~\ref{tides},
because it corresponds to
$\Delta t' = 90\,{\rm s}$,
$Q = 1300$,
$Q' = 44000$,
$E = 1.5\times 10^{-5}$.
Most importantly, though, this solution is no longer in contradiction
with the ECL dataset,
or the dynamics of the inner eclipsing binary.
This confirms that a five-component model is necessary to correctly describe
the $\xi$~Tau system.


\section{Temporal evolution of $\xi$~Tau} \label{evolution}

\subsection{Signs of secular evolution} \label{secular}

The $\xi$~Tau system is appreciably compact
to reveal temporal evolution of its individual orbits.
In particular, the observations available to-date
require the orbital elements of the eclipsing pair (Aa+Ab)
and of the third component (B)
to evolve according to our dynamical model
(Fig.~\ref{orbit1}).
These variations occur at various timescales,
(i)~the shortest (days to a year),
defined by the orbital periods $P_1$ and $P_2$;
(ii)~intermediate (decades to centuries),
defined by the orbital period $P_3$,
or the precession rates $\dot\Omega_1$, $\dot\Omega_2$, $\dot\omega_2$
expressed in the Laplace reference frame,
whose $z$-axis coincides withthe angular momentum
of the (Aa+Ab)+B triple;
(iii)~the longest (millenia to $10^5$ years),
given by the slow precession rate $\dot\Omega_3$.

\begin{figure}
\centering
\includegraphics[width=7.3cm]{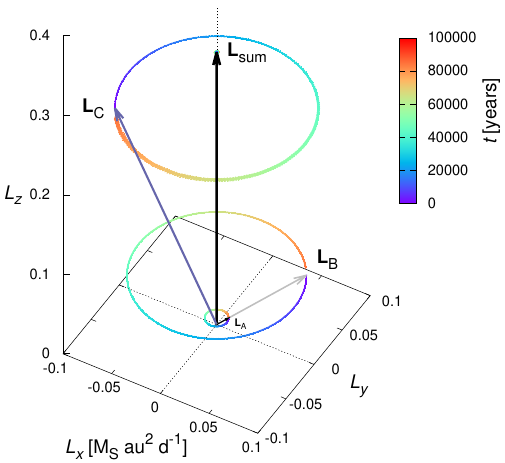}
\caption{
Angular momentum vectors of the $\xi$~Tau subsystems (Aa+Ab, B, C)
and their long-term evolution.
The coordinates were rotated to the Laplace reference frame,
with the z-axis along the total angular momentum $\vec L_{\rm sum}$
of the whole $\xi$~Tau system
(in absence of external torques $\vec L_{\rm sum}$ is constant).
The individual contributions were obtained as
$\vec L_{\rm A} = m_{\rm A}'\,\vec r_{\rm A}\times\dot{\vec r}_{\rm A}$
where $m_{\rm A}'$ denotes the reduced mass,
$\vec r_{\rm A}$ and $\dot{\vec r}_{\rm A}$
the Jacobi coordinate and velocity of the Aa+Ab binary
(and similarly for B and C parts,
$\vec L_{\rm B}$ and $\vec L_{\rm C}$).
The (2+1)+1 hierarchy of the system implies that
(i)~$\vec L_{\rm C}$ and
${\vec L}_{\rm AB} \equiv {\vec L}_{\rm A}+{\vec L}_{\rm B}$
regularly precess about $\vec L_{\rm sum}$
with a period of ${\simeq}85000$\,y
(see colours showing the time in the course of simulation),
preserving their mutual angle $J_2 \simeq 71^\circ$;
(ii)~$\vec L_{\rm A}$ and ${\vec L}_{\rm B}$
regularly precess about $\vec L_{\rm AB}$
with a period of ${\simeq}7000$\,d,
preserving their mutual angle $J_1 \simeq 0.5^\circ$.
Since $\vec L_{\rm C}$ holds the largest share of $\vec L_{\rm sum}$,
its angular distance from $\vec L_{\rm sum}$ is only $23^\circ$,
while that of ${\vec L}_{\rm AB}$ is more than $47^\circ$.
As a result, the observed inclinations with respected to the sky-plane,
$i_1$ and $i_2$,
exhibit large variations over millenia.
}
\label{l}
\end{figure}

\paragraph{Shortest timescales.}
The mutual gravitational interaction in the (Aa+Ab)+B triple
makes the shortest timescale apparent
in the periods $P_1$ and $P_2$
(or equivalently, the semimajor axes $a_1$ and $a_2$)
as well as the eccentricities $e_1$ and $e_2$.
We find that $e_1(t)$ oscillates between $0$ and $0.008$,
which is dominated by short-period terms with $P_1$ and $P_2$
periods and amplitudes $\simeq 0.003$ and $\simeq 0.004$.
These variations of $e_1$ are necessary
to explain the exact eclipse timings,
namely the contribution which cannot be produced
by the perturbation of binary's mean motion by star~B.


\paragraph{Intermediate timescales.}
The correlated variations of inclinations $i_1$ and~$i_2$,
as well as the nodal longitudes $\Omega_1$ and $\Omega_2$
in the observer's reference frame,
occur with the periodicity of $P_\Omega\simeq 7000$\,d.
The projected inclination $i_1$ of the inner eclipsing binary
changes from $86.1^\circ$ to $87.1^\circ$,
an effect clearly manifested in eclipse durations and depths,
according to the precise photometric measurements by
MOST (2012 and 2017) and TESS (from 2020 to 2023).
These variations appear to be a consequence of a simple precession
of the orbital angular momenta
$\vec L_{\rm A}$ and $\vec L_{\rm B}$
of the Aa+Ab binary and of the component B orbit
about their composite angular momentum
$\vec L_{\rm AB} = \vec L_{\rm A}+{\vec L}_{\rm B}$.
In particular, both 
$\vec L_{\rm A}$ and $\vec L_{\rm B}$
describe coni with small opening angles
about ${\vec L}_{\rm AB}$ with the period~$P_\Omega$,
preserving their mutual angle $J_1$ nearly constant.
Due to the very small value of~$e_1$,
the argument of periastron $\omega_1$ is an ill-defined orbital element
(especially near epochs when $e_1 \simeq 0$),
and its fast circulation is not a representative secular effect.
In contrast, the argument of periastron $\omega_2$ of orbit~B
exhibits in the observer's reference frame
a regular precession with a rate of
$2.1^\circ\,{\rm y}^{-1}$
\citep[see][and also Fig.~\ref{orbit1}]{Nemravova_2016A.A...594A..55N}.
This effect is evidenced by the measured RVs
of components Aa, Ab, and B. 

In order to explore the behavior of the orbital eccentricity $e_1$ of the
eclipsing binary Aa+Ab on intermediate timescales, we digitally filtered
the signal with periods $P_1$ and $P_2$ from the time series of osculating
values of $e_1$. The secular theory of triple dynamics 
\citep[e.g.,][]{Georgakarakos..2003MNRAS.345..340G,Georgakarakos..2009MNRAS.392.1253G,Breiter_2015MNRAS.449.1691B} predicts two contributions to $e_1$, namely
(i)~the free (proper) component, depending on the initial conditions, and
(ii)~the forced component, due to the presence of the third star (B)
and its perturbations.
We found that the forced component
with an amplitude $7.6\times 10^{-4}$ and a period of $167$\,yr
(i.e. the apsidal precession timescale of $\omega_2$)
dominates the evolution of $e_1$,
leaving no signal corresponding to the free component.
This result suggests that the free component of~$e_1$
has been damped by the past tidal evolution in the binary Aa+Ab,
a process that is currently witnessed by the weak tidal signal
(Sect.~\ref{tides}).

\paragraph{Longest timescales.}
The variations on the longest timescales are associated with
the outer orbit of component~C ($P_3 \doteq 18900\,{\rm d} \doteq 52\,{\rm y}$).
The gravitational perturbation of orbit~B
during the orbital motion of~C
is clearly manifested in the measured RVs.
It is amplified near the periastron passage of~C in July 2008
due to its large eccentricity, $e_3 \simeq 0.573$.
An interesting `sawtooth' structure seen during the sudden increase of~$i_2$
between JD~2454000 and 2455000
(an interval of time extending approximately 1.5\,y
before and after the periastron passage of~C;
Fig.~\ref{orbit1})
may be well-explained as a coherent effect during 5 or~6 conjunctions
between components~B and~C,
repeating themselves after 145\,d.
We verified that the conjunctions occur at the same geometrical configuration,
implying the out-of-plane perturbing acceleration~$W$ in orbit B
maintains the same sign.
Consequently, since
${\rm d}i/{\rm d}t \propto W \cos(f+\omega)$
\citep{Gauss_1809tmcc.book.....G},
the perturbing effect on $i_2$ accumulates coherently.

\begin{figure}
\centering
\includegraphics[width=7.3cm]{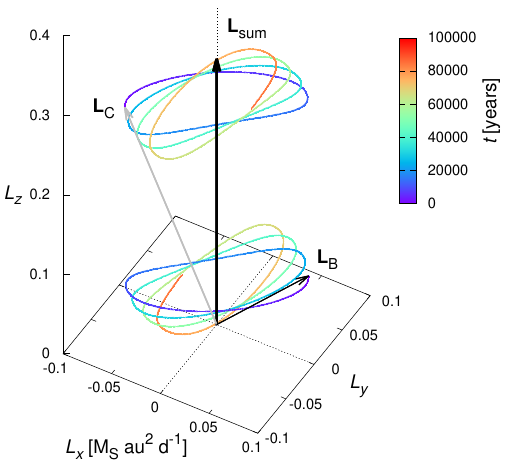}
\caption{
The same as Fig.~\ref{l}, but only for 3 bodies (A, B, C),
where A~represents Aa+Ab.
Without a protection mechanism,
the Kozai oscillations were triggered,
with a period of ${\simeq}11500\,{\rm y}$.
The angular momenta
${\vec L}_{\rm B}$ and ${\vec L}_{\rm C}$
still describe conical surfaces,
but now their mutual angle $J$ exhibits large oscillations,
which triggers correlated oscillation of eccentricities and inclinations
of the inner and outer orbits.
}
\label{l_3body}
\end{figure}


\subsection{Stability} \label{stability}

The gravitational perturbations operate also on timescales
longer than centuries,
beyond what is evidenced by the available observations.
We thus conducted numerical integrations over $10^5\,{\rm y}$
to study the stability of the $\xi$~Tau system.
We assumed the four-component model from Sect.~\ref{3.2}.
The integrator was the same,
the Bulirsch--Stoer with an adaptive time step,
but we used a smaller tolerance,
$\epsilon = 10^{-10}$,
to assure the angular momentum conservation.

We found that the ((Aa+Ab)+B)+C hierarchy is stable,
with all components moving in a well-orchestrated manner.
Apart from the (Aa+Ab)+B inner triple,
where $\vec L_{\rm A}$ and ${\vec L}_{\rm B}$
precess about ${\vec L}_{\rm AB}$,
as discussed in Sect.~\ref{secular},
we note that the nodal period~$P_\Omega$
is smaller than~$P_3$ (and non-resonant).
This implies that the corresponding nodal period
at the higher level in the $\xi$~Tau hierarchy,
namely the precession of
${\vec L}_{\rm AB}$ and ${\vec L}_{\rm C}$ about $\vec L_{\rm sum}$
is much longer,
${\simeq}85000$\,y
according to our model.

\paragraph{Non-existent Kozai oscillations.}

For comparison, we tested a 3\hyphenation{-}{-}{-}body model of $\xi$~Tau,
where the Aa+Ab binary was replaced with a~single component~A.
This totally changed the overall orbital evolution,
because of the Kozai oscillations
\citep{vonZeipel_1910AN....183..345V,Lidov_1961,Kozai_1962AJ.....67..591K}.
As illustrated in Fig.~\ref{l_3body},
the eccentricities $e_2$, $e_3$ and inclinations $i_2$, $i_3$
exhibit coupled oscillations,
with a period of ${\simeq}11500$\,y.
The peak value of $e_2(t)$ is $0.8$,
which would destabilize the innermost binary Aa+Ab.
Nevertheless, the stability of the (Aa+Ab)+B sub-system is ensured
by a protection mechanism,
consisting of its own mutual interaction,
which induces the periastron precession $\dot\omega_2$
with a period of only $180$\,y.
This is much shorter than the Kozai timescale,
effectively switching off the Kozai mechanism.
\vskip\baselineskip


\begin{figure*}
\centering
\begin{tabular}{@{}c@{}c@{}c@{}}
\includegraphics[width=6.0cm]{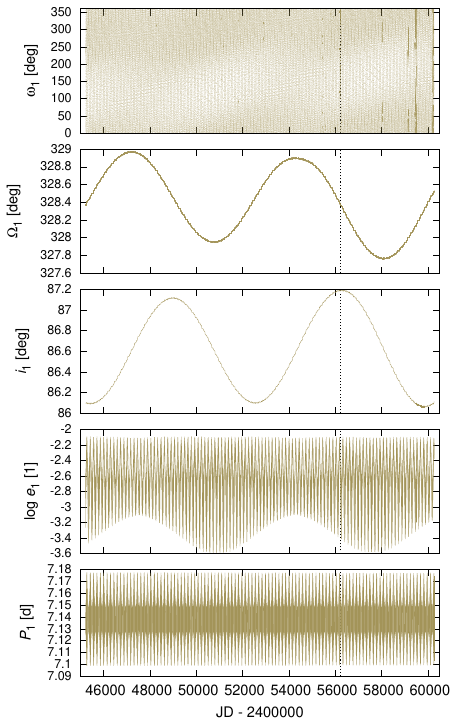} &
\includegraphics[width=6.0cm]{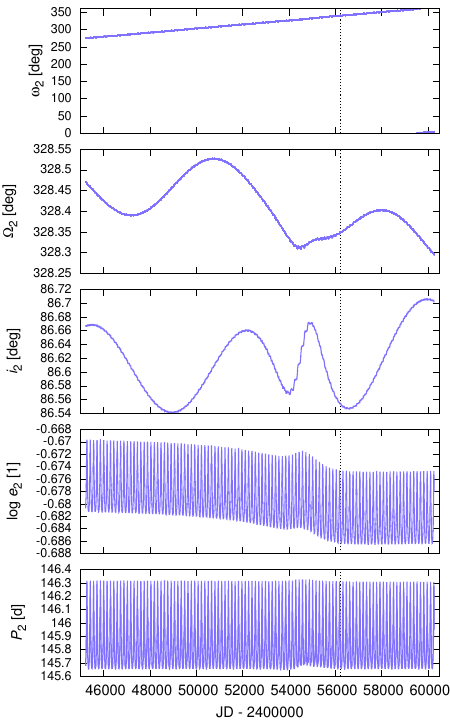} &
\includegraphics[width=6.0cm]{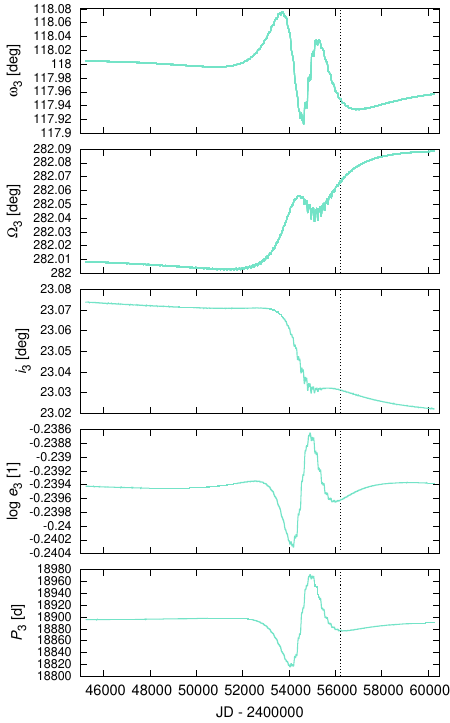} \\
\end{tabular}
\caption{
Osculating orbital elements of $\xi$~Tau in the course of time,
for the model with $\chi^2 = 3639180$.
The angular elements are expressed in the observer's reference frame.
See the description in the main text.
}
\label{orbit1}
\end{figure*}

\definecolor{cyan}{rgb}{0, 1, 1}
\definecolor{lightcyan}{rgb}{0.66, 1, 1}
\definecolor{lightgray}{rgb}{0.85, 0.85, 0.85}

\begin{figure*}[p]
\centering
\includegraphics[width=16.5cm]{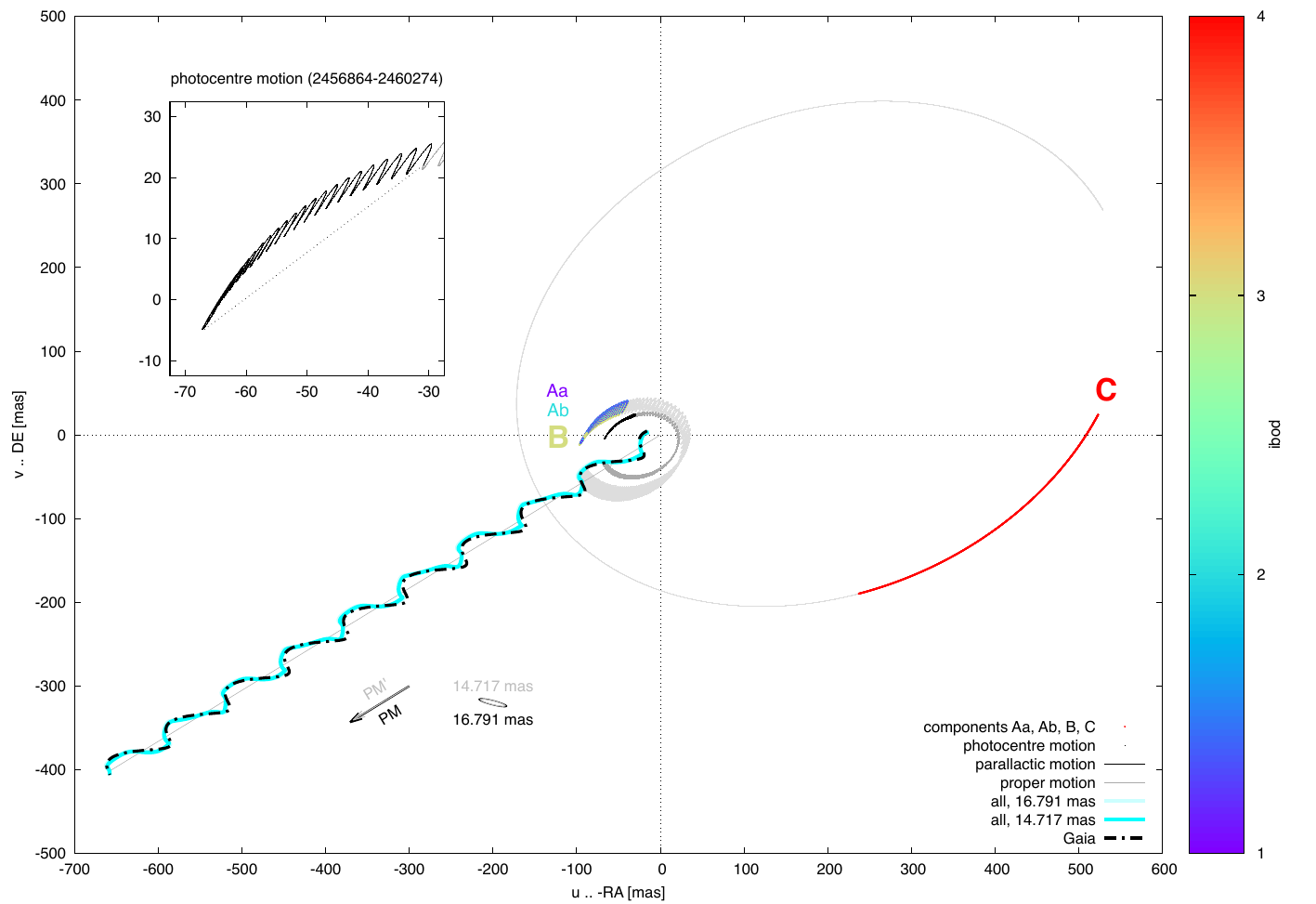}
\caption{
Photocentre motion of the $\xi$~Tau system
and its influence on parallax measuments.
We plot the orbital motion of individual components Aa, Ab, B, C
in the barycentric frame
(\textcolor{lightgray}{light gray} or coloured,
if within the observational timespan of Gaia),
the photocentre motion (black),
the parallactic motion (\textcolor{gray}{gray}),
the proper motion (\textcolor{lightgray}{light gray}), and
all contributions with the new parallax
$\pi' = 14.717\,{\rm mas}$
(\textcolor{cyan}{cyan}).
For comparison, we plot
all contributions with the old parallax
$\pi = 16.791\,{\rm mas}$
(\textcolor{lightcyan}{light cyan})
and also the old trajectory from Gaia DR3
({\bf dash-dotted}).
}
\label{photocentre2}
\end{figure*}


An interesting implication of the fact that
${\vec L}_{\rm C} \gg {\vec L}_{\rm AB}$
(see Fig.~\ref{l})
is the opening angle of the cone described by ${\vec L}_{\rm AB}$
is relatively large ($47^\circ$).
As a result, the viewing geometry of the eclipsing binary (Aa+Ab)
changes significantly over millennia.
We predict eclipses will vanish in 18000\,y from now.


\section{Discussion} \label{discussion}

\subsection{Photocentre motion in Gaia DR4} \label{gaia}

In order to understand how the multiplicity of $\xi$~Tau
influences its parallax measurements
\citep{Vallenari_2023A&A...674A...1G},
we computed its photocentre motion.
The simplest, 5-parameter model suitable for single stars consists of
the position $\alpha$, $\delta$,
the proper motion (PM) $\mu_\alpha$, $\mu_\delta$, and
the parallax $\pi$.
We added to this model the photocentre,
computed as an average of barycentric positions,
weighted by the passband fluxes of components Aa, Ab, B, C.
For reference,
the Gaia~G filter has
$\lambda_{\rm eff} = 639.02\,{\rm nm}$,
$\Delta_{\rm eff} = 317.32\,{\rm nm}$
and the respective weights were:
$0.181$,
$0.173$,
$0.603$, and
$0.042$.
Our results (Fig.~\ref{photocentre2}) show that
the overall photocentre motion
within the observational time span of Gaia
reaches almost $40\,{\rm mas}$.
However, this can be compensated by using different PM values,
specifically,
$\mu_\alpha' = 67.0\,{\rm mas}\,{\rm y}^{-1}$,
$\mu_\delta' = -40.5\,{\rm mas}\,{\rm y}^{-1}$.
In other words,
the new trajectory on the sky was adjusted
so that its end points correspond
to the old trajectory.

Now the photocentre motion consists of two contributions,
(i)~6\,mas oscillating due to the triple (Aa+Ab)+B;
(ii)~4\,mas curved due to component~C.
These offsets were most likely responsible
for an incorrect value of
$\pi = 16.791\,{\rm mas}$
in Gaia DR3.
When we used our preferred distance (67.9\,pc),
corresponding to $\pi' = 14.717\,{\rm mas}$,
the new trajectory became closer to the old trajectory,
or the actual measurements.
Consequently, our model explains not only all kinds of observations,
but also why the Gaia DR3 parallax is offset.
We do expect that binary, triple or quadruple astrometric models
in Gaia DR4 will confirm our results.




\subsection{Oscillations} \label{oscillations}

\begin{figure}[t]
\centering
\includegraphics[width=8.5cm]{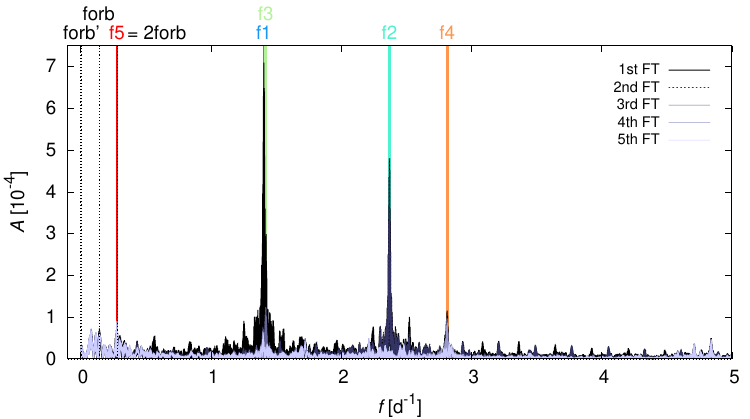}
\includegraphics[width=8.5cm]{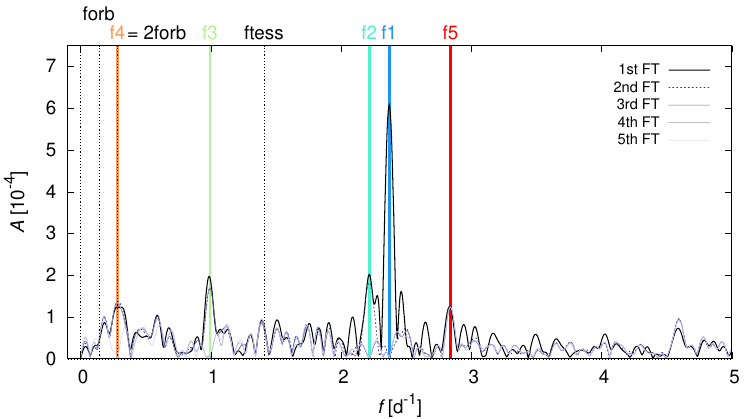}
\caption{
Periodogram of $\xi$~Tau from the TESS (top) and MOST 2012 (bottom) light curves.
We removed eclipses and computed five Fourier transforms
\citep{Lenz_2005CoAst.146...53L},
with fitting of frequencies, amplitudes and phases,
and subtracting the respective model,
$\sum_i A_i \sin[2\pi(f_i t + \phi_i)]$.
For TESS, the dominant frequencies were
1.407,
2.371,
1.421,
2.814, and
$0.279\,{\rm d}^{-1}$.
We note $f_4 = 2f_1$ and $f_5 = 2f_{\rm orb}$ of the inner orbit.
For MOST 2012, on the contrary, we found
2.369,
2.216,
0.993,
0.282, and
$2.839\,{\rm d}^{-1}$.
The non-existent $1.407\,{\rm d}^{-1}$ frequency is labeled ``ftess''.
}
\label{xitau_fou}
\end{figure}

In order to better understand rapid low-amplitude variability of $\xi$~Tau
\citep{Nemravova_2016A.A...594A..55N},
we used the new photometric data from TESS
to compute a new periodogram.
After excluding all eclipses intervals
to suppress the orbital frequency $f_{\rm orb}$ (and its harmonics),
we actually computed a sequence of periodograms
\citep{Lenz_2005CoAst.146...53L},
revealing two dominant frequencies
$f_1 = 1.407\,{\rm d}^{-1}$,
$f_2 = 2.371\,{\rm d}^{-1}$,
with amplitudes approximately $7$ and $5\times 10^{-4}$ in the relative flux
(Fig.~\ref{xitau_fou}).
The corresponding periods are
$0.710$ and $0.421\,{\rm d}$,
respectively.
The overall time span (1115\,d)
allows us to reach the resolution of $1/\Delta = 10^{-4}\,{\rm d}^{-1}$.

The $f_2$ frequency is persistent and
could be attributed to rotation of the 3rd component
\citep{Nemravova_2016A.A...594A..55N}.
Its rotational broadening
$v_{\rm rot3} = 242\,{\rm km}\,{\rm s}^{-1}$
corresponds to the upper limit of
$P_{\rm rot3} \le 0.472\,{\rm d}$.
The rotational axis inclination is unknown,
but in order to obtain 0.421\,d, it should be ${\sim}60^\circ$.

The $f_1$ frequency should be also attributed to the 3rd component,
which is most likely a slowly-pulsating B star
(SPB; \citealt{Waelkens_1985A&A...152....6W}).
The star is located in the instability strip of g-modes
of the order $\ell = 1$, or 2
\citep{Sharma_2022MNRAS.515..828S}.
However, there is a problem with non-existent $f_1$ oscillations
in the MOST~2012 data,
and no evidence of damping in the TESS 2020--2023 data. 
A possible explanation may consist of interference (`beat')
of two nearby frequencies
(e.g., \citealt{Sharma_2022MNRAS.515..828S}, Fig.~5).
However, the time resolution of the TESS photometry is insufficient
to identify frequencies ($f_1$, $f_1'$) whose combination would result
in conjectured beats ($f_1-f_1'$)
separated by more than 8~years.

We verified that the reason for missing $f_1$ oscillations is not technical.
The MOST telescope is a Maksutov--Cassegrain
with a Fabry lens placed in the secondary focus
\citep{Walker_2003PASP..115.1023W}.
Its diaphragm (0.117\,mm)
corresponds to the angular diameter of $\phi = d/f = 27''$,
which is too large to miss light from the 4th component
(up to 600\,mas).

Still on a technical note,
the TESS satellite observed $\xi$~Tau as an over-exposed target,
with blooming
\citep{tess,Krishnamurthy_2019AcAau.160...46K}.
In order to compute a sum of all pixels (both over- and under-exposed)
the photometric aperture must be
13 pixels in $y$-direction.
Since the pixel size is $15\,\mu{\rm m}$,
the focal length 146\,mm,
the angular distance is up to $\phi = 4.2'$.
Using Lightkurve
\citep{Cardoso_2018ascl.soft12013L},
we checked stars in the surroundings.
In particular,
TIC 399947175
is close to the aperture,
but does not exhibit such oscillations.
So, unfortunately, the $f_1$ oscillations remain an open problem.

\section{Conclusions} \label{conclusions}

In this work, we studied the $\xi$~Tau quadruple system
using dynamical models of differing complexity,
constrained by 11 different types of observations.
Using N-body and post-Newtonian terms
and assuming four components (Aa, Ab, B, C),
we found a global minimum of $\chi^2$, where
the total mass is $9.55\,M_\odot$ and
the distance $66.1\,{\rm pc}$
(cf. Table~\ref{params}).
However, such model still exhibits some tension, for example,
minima timings for the MOST 2017 dataset are offset by ${\sim}0.0015\,{\rm d}$,
the SED is overestimated by 30\,\%,
etc.

It was surprisingly difficult to find a better solution.
Eventually, we were forced to use
N-body, post-Newtonian, oblateness, and tides terms
and assume five components (Aa, Ab, B, Ca, Cb).
We found a global minimum, where
the total mass is $10.37\,M_\odot$,
mainly due to the binary nature of component C,
and the distance $67.9\,{\rm pc}$.
Given the increased number of model parameters
(36 vs. 44),
it is not surprising that the model fits observations better.
Nevertheless, the sole behavior of the model,
how the minima timings were corrected,
how the SED was corrected,
suggests that the five-component model is the correct one.

Finally, let us point out that $\xi$~Tau is indeed an interacting stellar system;
even the most distant component~C,
interacts with component B at periastron passage,
inducing a non-negligible phase shift,
which propagates to the innermost, eclipsing binary Aa+Ab,
shifting measurably its eclipse timings.
This has a potential to detect additional,
dwarf or exoplanetary components with low masses.

\begin{acknowledgements}
This work has been supported by the Czech Science Foundation through the grant
25-16507S (M.~Bro\v z and D.~Vokrouhlick\'y).
We thank an anonymous referee for constructive comments.
This work is co-funded by the European Union (ERC, MAGNIFY, Project 101126182).
Views and opinions expressed are however those of the authors only and do not necessarily
reflect those of the European Union or the European Research Council.
Neither the European Union nor the granting authority can be held responsible for them.
This work has made use of data from the European Space Agency (ESA) mission Gaia,
processed by the Gaia Data Processing and Analysis Consortium (DPAC).
This research has made use of the SIMBAD database operated at CDS, Strasbourg (France),
and NASA’s Astrophysics Data System (ADS).
This paper includes data collected by the TESS mission,
which are publicly available from the Mikulski Archive for Space Telescopes (MAST).
Funding for the TESS mission is provided by NASA's Science Mission directorate.
\end{acknowledgements}


\bibliographystyle{aa}
\bibliography{references}


\clearpage
\appendix

\section{Rectification of CTIO/CHIRON spectra}\label{rectification}

As a prerequisite,
one has to describe the CTIO/CHIRON instrument,
in order to rectify all observed spectra
and minimize systematics,
in particular, in the Balmer lines wings.
This required redefinition of the blaze function.
The rectification is defined as
\begin{equation}
I_i^{\rm rect}(\lambda) \equiv {I_i^{\rm obs}(\lambda)\over R(\lambda)}\,,\label{eq1}
\end{equation}
where
$I_i^{\rm rect}(\lambda)$ is the rectified spectrum,
$I_i^{\rm obs}(\lambda)$ the observed spectrum,
$R(\lambda)$ the rectification function;
in the case of an echelle spectrum,
it is called the `blaze function'.
The motivation for rectification is
to avoid absolute calibration,
to measure line profiles, and
to compare to normalized synthetic spectra.
For complex echelle spectra,
a number of methods exist:
spline fitting,
using a reference spectrum,
a theoretical blaze function,
$\alpha$-shapes \citep{Xu_2019AJ....157..243X}, or
optimal extraction \citep{Piskunov_2021A&A...646A..32P}.

In the case of spline fitting
(e.g. \citealt{Harmanec_2020A&A...639A..32H}),
parts considered as continuum are fitted with a cubic spline,
where the spline extends over wide lines.
Among the common problems is 
uncertain continuum level,
uncertain line wings, or
unreliable $\log g$ values.
Moreover, in the Paschen series region,
overlapping lines and missing continuum
make reliable rectification impossible.

In the case of a reference spectrum
\citep{Skoda_2008SPIE.7014E..5XS},
a rectified spectrum of a similar star is used.
Alternatively, it could be a spectrum rectified by splines,
obtained by comparison to a suitable synthetic spectrum.
A measured spectrum is then divided by a reference spectrum
(Eq.~(\ref{eq1})),
to get $R(\lambda)$.
Of course, this cannot be done without a low-noise reference spectrum.
Among the common problems is
noise in the reference spectrum,
mismatch of resolution, or
rapid changes.

Rectification by a theoretical blaze function
\citep{Barker_1984AJ.....89..899B}
assumes that
\begin{equation}
R(\lambda) =  A_r \left\{{\sin\left[\pi\alpha_r X_r(\lambda)\right] \over \pi\alpha_r X_r(\lambda)}\right\}^2\quad\hbox{for }\forall r\,,\label{eq:R_r}
\end{equation}
where the coefficients are different for each order~$r$, and
\begin{equation}
X_r(\lambda) = r_{\rm e} \left(1 - {\lambda\over\lambda_{{\rm c},r}}\right)\,,\label{eq:X}
\end{equation}
where
$\lambda$ is the wavelength,
$\lambda_{\rm c}$ the central wavelength,
$A$ the local maximum of the blaze function,
$r_{\rm e}$ the echelle order;
we use our own numbering of orders
\begin{equation}
r \equiv 126 - r_{\rm e}\,.
\end{equation}
The grating constant
\begin{equation}
\alpha = {l\over d}\cos\gamma\,,     
\end{equation}
where
$l$ is the width of a grating facet,
$1/d$ the spatial frequency of facets,
$\gamma$ the blaze angle.

Unfortunately, Eq.~(\ref{eq:R_r}) only accounts for the basic grating theory.
Slight variations are not included
and it does not represent the actual CTIO/CHIRON data.
The blaze function thus needs more degrees of freedom
and our method is based on this approach.

\begin{figure}
\centering
\includegraphics[width=8.5cm]{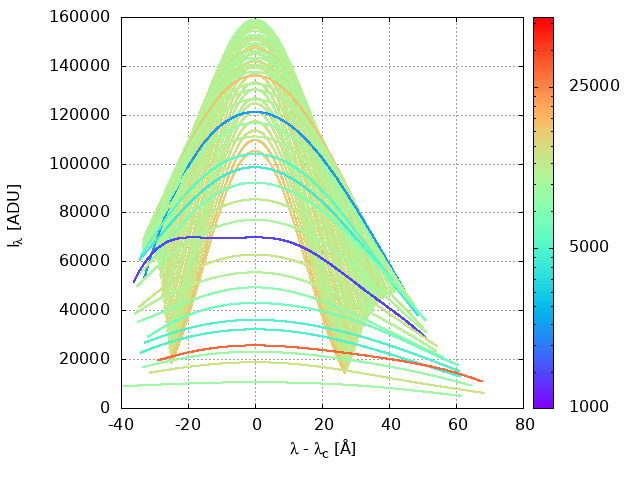}
\caption{
Redefined blaze function $R(\lambda)$,
according to (Eq.~(\ref{R_lambda})),
estimated for CTIO/CHIRON spectra.
It is plotted in absolute, analog-digital units (adu) and
for shifted wavelengths $\lambda-\lambda_{{\rm c},r}$,
for each order~$r$.
The rectification method was resimplex1.
Colours correspond to $\chi^2$ contributions.
The total $\chi^2 = 666239$ (unreduced)
is a measure of correspondence between
the rectified observed spectra
and the synthetic (rectified) spectra.
}
\label{blaze_chi2_avari_rs1}
\end{figure}


\subsection{Redefined blaze function} \label{4.1}

We proposed the following redefinition of the blaze function
\begin{equation}
R(\lambda) = A_r \left\{ {\sin\left[\pi\beta_r(\lambda)X_r(\lambda)\right]\over \pi\beta_r(\lambda)X_r(\lambda)} \right\}^2 \quad\hbox{for }r = 1..62\,, \label{R_lambda}
\end{equation}
where the constant $\alpha_r$ is replaced with the function
\begin{eqnarray}
\beta_r(\lambda) &=&
    10^{-10} \gamma_r(\lambda_{{\rm c},r} \!-\! \varepsilon_r \!-\! \lambda)^4
+   10^{-8}  \theta_r(\lambda_{{\rm c},r} \!-\! \varepsilon_r \!-\! \lambda)^3 + \nonumber\\
&+& 10^{-6}  \delta_r(\lambda_{{\rm c},r} \!-\! \varepsilon_r \!-\! \lambda)^2
+   \alpha_r\,,
\label{eq:beta}
\end{eqnarray}
while $X_r(\lambda)$ remained the same as in Eq.~(\ref{eq:X});
$\lambda_{{\rm c},r}$, $A_r, \alpha_r, \delta_r, \gamma_r, \theta_r, \varepsilon_r$
are order-dependent parameters,
which we determined by fitting.

We defined the respective metric as
\begin{equation}
\chi^2 \equiv \sum_{i=1}^N\left({I_i^{\rm rect} - I_i^{\rm syn}\over\sigma_i}\right)^2\,,
\end{equation}
where $I_i^{\rm rect}$ is the rectified observed intensity,
$I_i^{\rm syn}$ is the intensity of the synthetic (rectified) spectrum;
the uncertainties
\begin{equation}
\sigma_i = \sqrt{\eta} {I_i^{\rm rect}\over\sqrt{I_i^{\rm obs}}}\,,
\end{equation}
where $I_i^{\rm obs}$ is the observed (not rectified) intensity,
$\eta$ is the gain of the CCD detector.

Synthetic spectra were taken from our previous, converged model of $\xi$~Tau
\citep{Nemravova_2016A.A...594A..55N}.
They served primarily to estimate the continuum level,
not the depths of lines.
We avoided the telluric lines,
as obtained by the Skycalc tool
\citep{Noll_2012A&A...543A..92N,Jones_2013A&A...560A..91J}
for the nearest available observatory,
which is La Silla.
The positions of the respective lines were masked in observed spectra.

In total, one has
2 parameters per-order and per-dataset 
($\lambda_{{\rm c},r}, A_r$)
and 5 parameters per-order
($\alpha_r, \delta_r, \gamma_r, \theta_r, \varepsilon_r$),
the number of orders is 62,
the number of datasets is 10
(since we selected a subset, evenly distributed in time),
which implied 1550 parameters.
We optimized their values by the simplex algorithm
\citep{Nelder_Mead_1965},
order by order.
We call this first, direct method `resimplex1'.%
\footnote{\url{http://sirrah.troja.mff.cuni.cz/~mira/xitau/}}
The resulting blaze function
is shown in Fig.~\ref{blaze_chi2_avari_rs1},
and it is significantly better than our own, manual rectification.


\subsection{Constrained blaze function} \label{4.2}

Alternatively,
in order to decrease the variability of the blaze function,
some parameters were approximated by polynomials
\begin{eqnarray}
\alpha(r)      &=& a_1 + a_2 r + a_3 r^2 + a_4 r^3 + a_5 r^4\,, \label{alpha_r}\\
\gamma(r)      &=& g_1 + g_2 r + g_3 r^2 + g_4 r^3 + g_5 r^4\,, \label{gamma_r}\\
\delta(r)      &=& d_1 + d_2 r + d_3 r^2 + d_4 r^3 + d_5 r^4\,, \label{delta_r}\\
\theta(r)      &=& t_1 + t_2 r + t_3 r^2 + t_4 r^3 + t_5 r^4\,, \label{theta_r}\\
\varepsilon(r) &=& e_1 + e_2 r + e_3 r^2 + e_4 r^3 + e_5 r^4\,, \label{epsilon_r}
\end{eqnarray}
while
$R(\lambda)$ and $\beta_r(\lambda)$ functions remained the same.
In the 1st step,
the respective coefficients were estimated by fitting these polynomials to
$\alpha_r$, $\gamma_r$, $\delta_r$, $\theta_r$, $\varepsilon_r$ discrete values
(Fig.~\ref{alpha(order)}).
The remaining parameters
$\lambda_{{\rm c},r}$, $A_r$
were taken from Sect.~\ref{4.1}.
In the 2nd step,
we further optimized
$a_1, a_2, \dots, e_4, e_5$,
and
$\lambda_{{\rm c},r}$, $A_r$
were periodically updated during convergence,
in order to improve their consistency with the polynomials.
We again worked with
62 orders,
10 datasets,
which implied 1265 parameters.
We call this second, constrained method `resimplex2'.

In order to avoid problems with wide lines,
we did not use the following orders for fitting (in $\AA$):
\#7 4732.3--4786.8 H$\beta$, 
\#8 4772.3--4827.3 H$\beta$, 
\#9 4813.1--4868.5 H$\beta$,
\#10 4854.5--4910.5 H$\beta$,
\#11 4896.7--4953.2 H$\beta$,
\#39 6472.5--6547.0 H$\alpha$,
\#40 6547.7--6623.1 H$\alpha$,
\#44 6867.1--6946.1 telluric lines,
\#52 7609.5--7697.0 telluric lines,
\#57 8160.9--8254.7 telluric lines.
The polynomials (Eqs.~(\ref{alpha_r})--(\ref{epsilon_r})) then serve as interpolants for these regions.

\begin{figure}
\centering
\includegraphics[width=8.5cm]{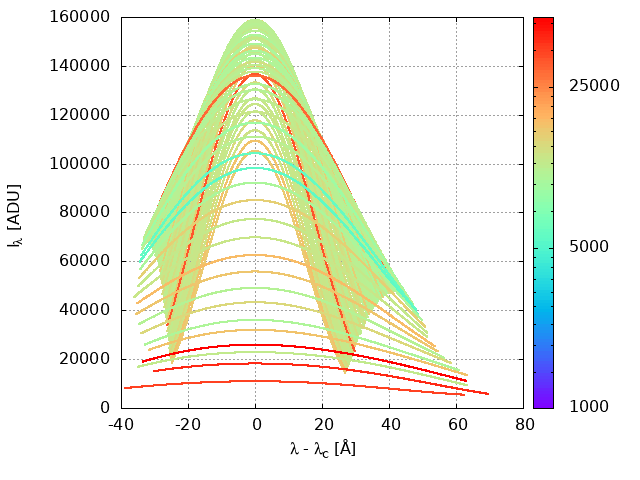}
\caption{
Same as Fig.~\ref{blaze_chi2_avari_rs1},
but for the rectification method resimplex2.
The respective $\chi^2 = 891923$.
}
\label{blaze_chi2_avari_rs2}
\end{figure}

The result is shown in
Fig.~\ref{blaze_chi2_avari_rs2}.
The $\chi^2$ is higher,
but still better than our manual rectification.
The biggest contribution is from the last four orders
(i.e. the Paschen series).

\subsection{Constrained central wavelengths} \label{4.3}

In order to further decrease the order-to-order variability,
we constrained the central wavelength as
\begin{equation}
\lambda_{\rm c}(r) = \sum_{i=1}^{10} l_i r^{i-1}
\label{lambda_c}
\end{equation}
and also the amplitude as 
\begin{equation}
A(r) = \sum_{i=1}^{9} a_i r^{i-1}\,.
\end{equation}
This turned out to be the compromise between simplicity and complexity,
keeping l`  ow $\chi^2$.
However, both $\lambda_{\rm c}(r)$ and $A_r(r)$ must be different
for each dataset,
because they are directly related to the variability of the atmosphere
(extinction, refraction).
Other blaze parameters were taken from Sect.~\ref{4.2}.
We call this third method `resimplex3'.
For technical reasons,
spectra had to be split at orders 16 (CCD amplifiers), 28, 48 (Paschen series).
We processed all 277 datasets this way,
but we worked with the datasets sequentially,
which limited the number of parameters.

\begin{figure}
\centering
\includegraphics[width=8.5cm]{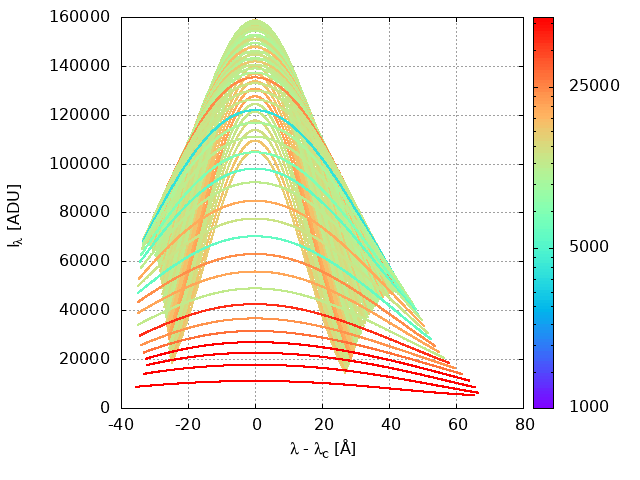}
\caption{
Same as Fig.~\ref{blaze_chi2_avari_rs1},
but for the rectification method resimplex3.
The respective $\chi^2 = 1217823$. 
}
\label{blaze_chi2_avari_rs3}
\end{figure}

The resulting blaze function is shown in
Fig.~\ref{blaze_chi2_avari_rs3}.
Specifically, how the rectification works in the H$\alpha$ line region is shown in
Fig.~\ref{synthetic_comps}.
The respective $\chi^2$ value is even higher,
but still acceptable,
if one avoids the Paschen series region.
In fact, the relatively higher $\chi^2$ value
does not imply worse rectification.
Instead, interpolating over some orders
(H$\alpha$, H$\beta$)
implies lower systematic errors due to the uncertain continuum level.

\clearpage

\onecolumn

\section{Supplementary figures}

\begin{figure*}[h!]
\vskip-\baselineskip
\centering
\includegraphics[width=6cm]{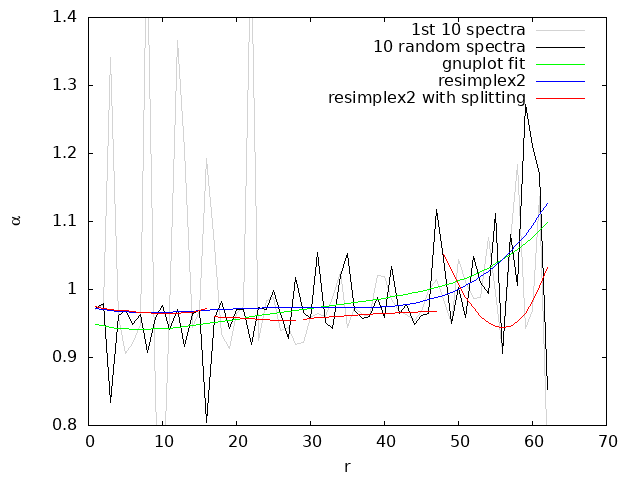}
\includegraphics[width=6cm]{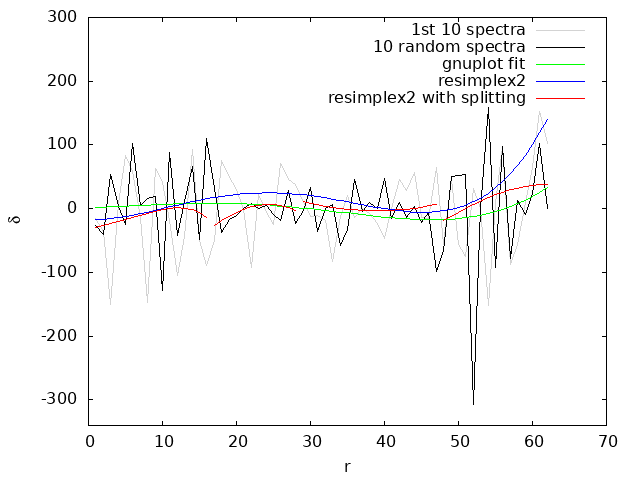}
\includegraphics[width=6cm]{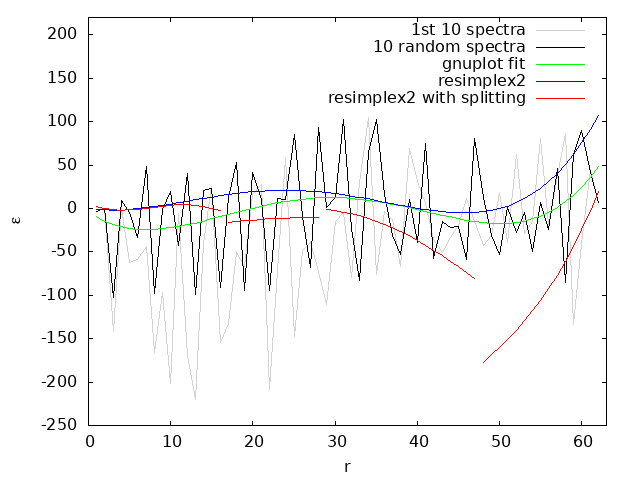}
\includegraphics[width=6cm]{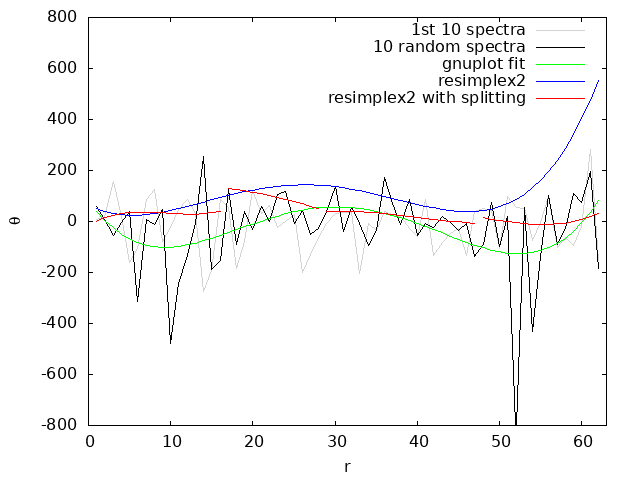}
\includegraphics[width=6cm]{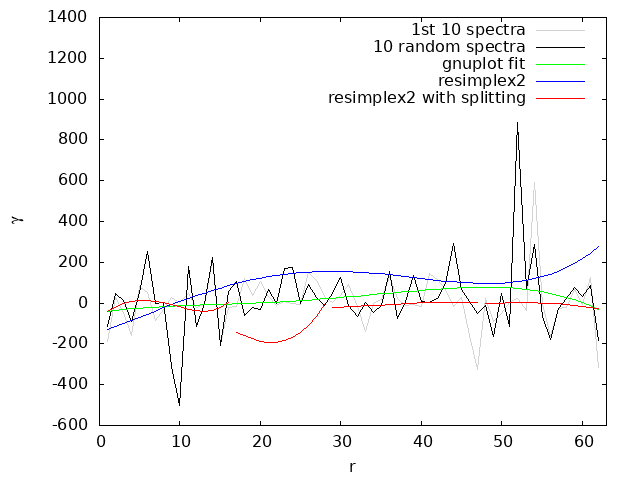}
\caption{
Parameters of the blaze function as a function of the order~$r$,
shown either as discrete values
$\alpha_r$, $\delta_r$, $\epsilon_r$, $\theta_r$, $\gamma_r$,
or as continuous polynomials
$\alpha(r)$, $\delta(r)$, $\epsilon(r)$, $\theta(r)$, $\gamma(r)$.
We show a comparison of two rectification methods:
resimplex1 on a set of 10 spectra ({\textcolor{black}{black}}),
resimplex1 on a different set of 10 spectra (\textcolor{gray}{gray}),
resimplex2 (\textcolor{blue}{blue}),
resimplex2 with splitting at orders 16, 28, 48 (\textcolor{red}{red}).
We also show a preliminary fit of black values (\textcolor{green}{green}),
which served as a starting for resimplex2.
The variability of discrete values is substantial,
especially for orders $r \gtrsim 50$ (Paschen series),
but this is suppressed when using polynomials.
}
\label{alpha(order)}
\end{figure*}

\begin{figure*}[h!]
\vskip-\baselineskip
\centering
\includegraphics[width=8cm]{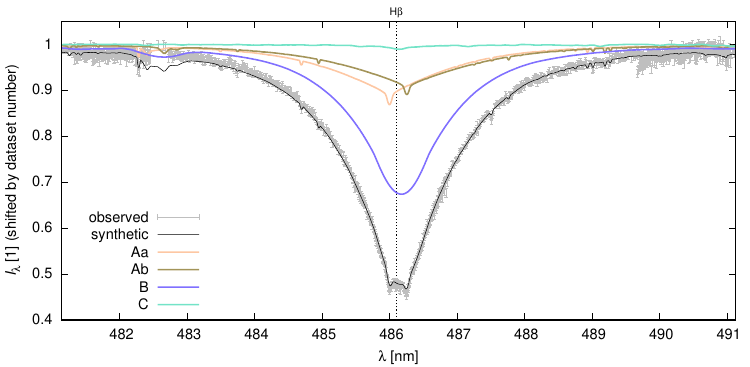}
\caption{
Synthetic spectra of individual components Aa, Ab, B, C
for the model with $\chi^2 = 3639180$,
in the H$\beta$ line region.
}
\label{synthetic2_4861}
\end{figure*}

\begin{figure*}[h!]
\vskip-\baselineskip
\centering
\includegraphics[width=8.5cm]{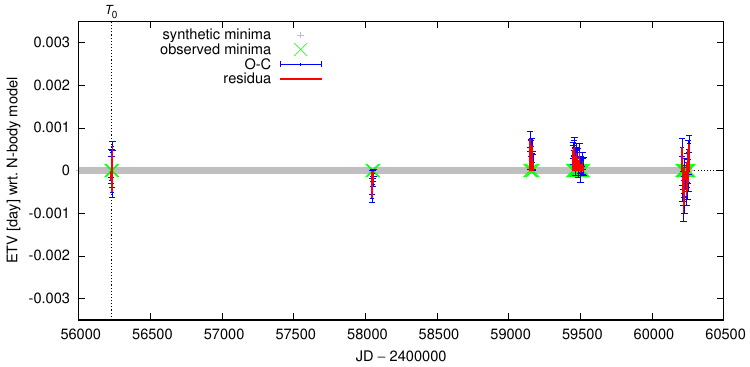}
\caption{
Same as Fig.~\ref{reference}
for non-zero tides,
with revised uncertainties of ETVs
of the order of $10^{-4}\,{\rm d}$.
The total $\chi^2 = 4145$.
Systematics of ETVs were largely compensated
by suitable a value of $\Delta t$.
}
\label{fitting22_allrvs_SINGLE__4145}
\end{figure*}

\begin{figure*}[p]
\centering
\includegraphics[width=19.1cm]{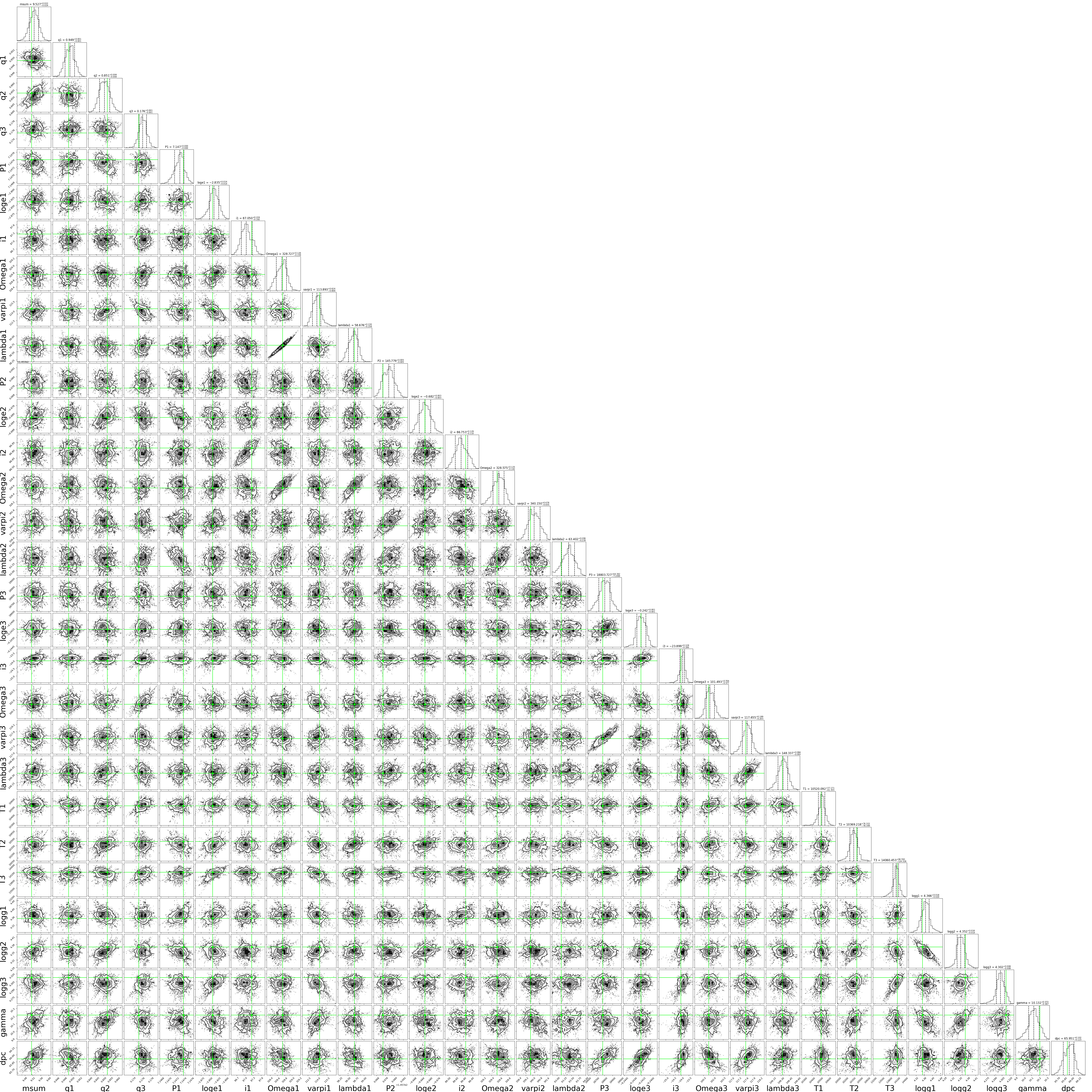}
\caption{
MCMC analysis
for the reference model with $\chi^2 = 1553$,
constrained by SKY, RV, ETV, ECL, SED, SED2 datasets.
The corner plot shows uncertainties and correlations of 30 parameters,
which were sampled with 64 walkers,
1000 iterations,
with 500 burn-in steps.
The highest-probability model is plotted (\color{green}green\color{black}).
Specifically,
$m_{\rm sum} = 9.572\pm 0.023\,M_\odot$,
$q_1 = 0.949\pm 0.001$,
$q_2 = 0.851\pm 0.004$,
$q_3 = 0.176\pm 0.001$,
$P_1       =  7.14730\pm 0.00010\,{\rm d}$,
$\log e_1  =  -2.835\pm 0.014$,
$i_1       =  87.050\pm 0.204^\circ$,
$\Omega_1  = 328.727\pm 0.143^\circ$,
$\varpi_1  = 113.833\pm 0.646^\circ$,
$\lambda_1 =  58.676\pm 0.144^\circ$,
$P_2       = 145.779\pm 0.005\,{\rm d}$,
$\log e_2  =  -0.682\pm 0.003$,
$i_2       =  86.753\pm 0.120^\circ$,
$\Omega_2  = 328.575\pm 0.133^\circ$,
$\varpi_2  = 340.150\pm 0.666^\circ$,
$\lambda_2 =  63.402\pm 0.223^\circ$,
$P_3       = 18803.7\pm 68.1\,{\rm d}$,
$\log e_3  =  -0.242\pm 0.001$,
$i_3       = -23.899\pm 0.157^\circ$,
$\Omega_3  = 101.493\pm 0.642^\circ$,
$\varpi_3  = 117.655\pm 0.196^\circ$,
$\lambda_3 = 148.337\pm 0.088^\circ$,
$T_1 = 10520\pm 39\,{\rm K}$,
$T_2 = 10369\pm 77\,{\rm K}$,
$T_3 = 14060\pm 107\,{\rm K}$,
$\log g_1 = 4.366\pm 0.019$,
$\log g_2 = 4.352\pm 0.015$,
$\log g_3 = 4.302\pm 0.008$,
$\gamma = 10.132\pm 0.070\,{\rm km}\,{\rm s}^{-1}$,
$d = 65.951\pm 0.112\,{\rm pc}.$
These 1-$\sigma$ uncertainties are considered local and small,
compared to global ones in Table~\ref{params},
because of some systematics and tension remained in the model.
}
\label{mcmc17_NOVIS_SINGLE}
\end{figure*}

Here we show
parameters of the blaze function (Fig.~\ref{alpha(order)}),
synthetic spectra of individual components Aa, Ab, B, C (Fig.~\ref{synthetic2_4861}),
the ETVs of $\xi$~Tau for the model with tides (Fig.~\ref{fitting22_allrvs_SINGLE__4145}), and
the MCMC analysis for the model without tides (Fig.~\ref{mcmc17_NOVIS_SINGLE}).

\begin{figure*}[p]
\centering
\leavevmode\kern0cm\includegraphics[width=18.5cm]{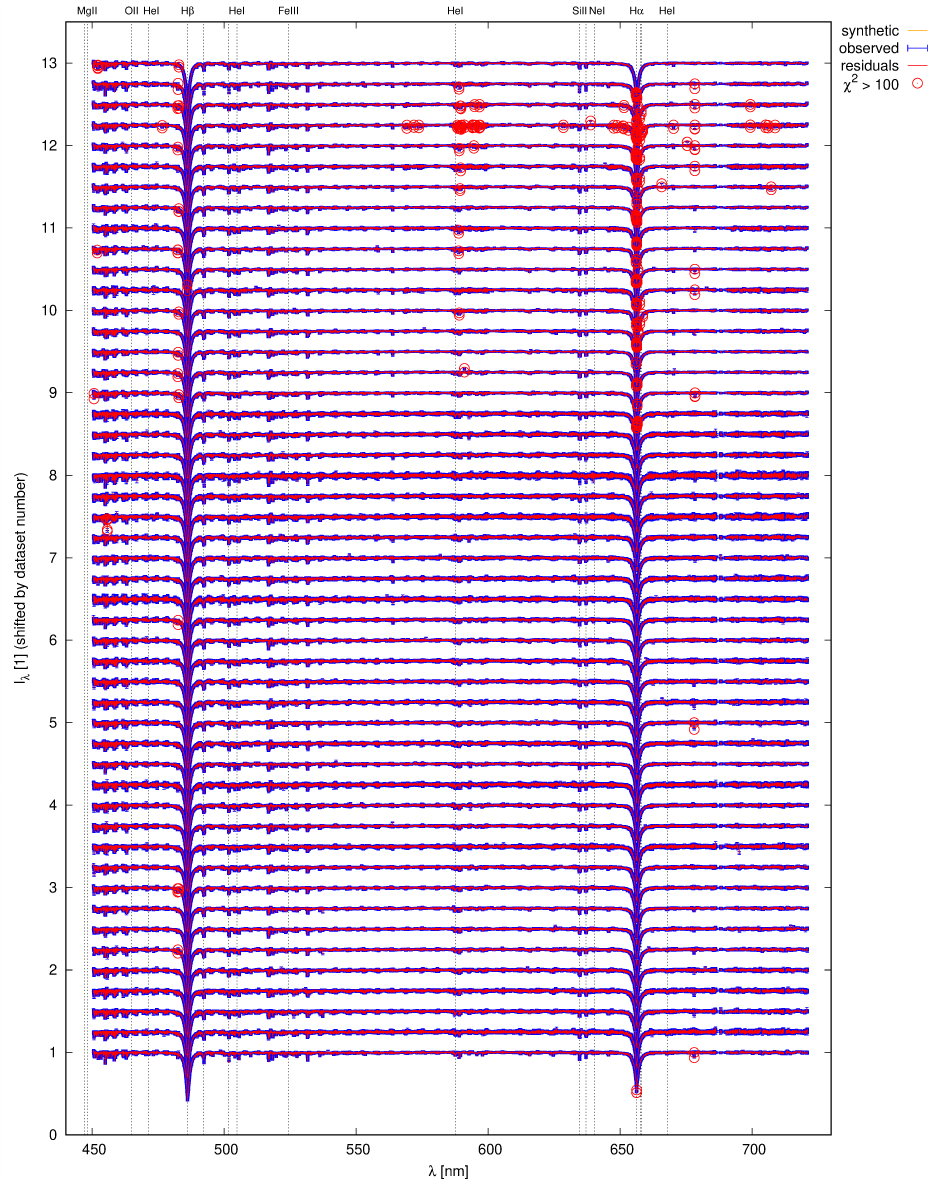}
\caption{
Comparison of CTIO/CHIRON observed spectra (one per night)
with synthetic spectra,
for the model with $\chi^2 = 3639180$.
The fit is acceptable, including the H$\alpha$, H$\beta$ wings,
with a notable exception of the H$\alpha$, H$\beta$ core depths
for Aa, Ab components
(see \textcolor{red}{red} circles for spectra
with higher signal-to-noise ratio).
}
\label{chi2_SYN}
\end{figure*}


\clearpage

\section{Supplementary tables}

Here we show
the journal of photometric observations (Table~\ref{jouphot}) and
dependent parameters of $\xi$~Tau models (Table~\ref{dependent})

\begin{table*}[h!]
\caption[]{Journal of photometric observations.}
\label{jouphot}
\centering
\tiny
\begin{tabular}{rcrccll}
\hline\noalign{\smallskip}
Station&Time interval& No. of &Passbands&HD of comparison&Source\\
       &(HJD$-$2400000)&obs.  &  &/ check star\\
\noalign{\smallskip}\hline
\hline\noalign{\smallskip}
 61&47909.64--48695.02&   70&\hp  &  --       &\citet{esa97}\\
  1&54116.25--54134.40&   35&\ubv &21686/21933& This paper\\
  1&54468.34--54720.57&   13&\ubv &21686/21933& This paper\\
  1&54862.31--54862.32&    3&\ubv &21686/21933& This paper\\
  1&55545.35--55545.36&    3&\ubv &21686/21933& This paper\\
  1&55574.25--55849.64&  238&\ubv &21686/21933& This paper\\
  1&55850.37--55850.54&    9&\ubv &21933/21686& This paper\\
  1&55851.40--55852.47&   21&\ubv &21686/21933& This paper\\
  1&55853.37--55850.50&   12&\ubv &21933/21686& This paper\\
  1&55854.49--55945.33&   97&\ubv &21686/21933& This paper\\ 
128&56222.00--56238.03&18510&MOST&    --      & \citet{Nemravova_2016A.A...594A..55N}\\ 
  1&56520.59--56527.60&    9&\ubvr&21686/21933& This paper\\
  1&56882.61--56882.63&    3&\ubvr&21686/21933& This paper\\
  1&57634.55--57656.64&    9&\ubvr&21686/21933& This paper\\
  1&57775.27--58079.35&   20&\ubvr&21686/21933& This paper\\
128&58044.20--58058.55&12179&MOST &  --       & This paper\\  
  1&58132.30--58402.48&   64&\ubvr&21686/21933& This paper\\
  1&58490.21--58490.32&   21&\ubvr&21686/21933& This paper\\
  1&58858.24--58865.42&   42&\ubvr&21686/21933& This paper\\
110&59144.53--59169.94& 1630&TESS &  --       & This paper\\
110&59447.71--59473.15& 1706&TESS &  --       & This paper\\
110&59474.18--59498.88& 1590&TESS &  --       & This paper\\
110&59500.20--59524.44& 1592&TESS &  --       & This paper\\
110&60208.55--60230.89& 2466&TESS &  --       & This paper\\
110&60237.00--60259.49& 2479&TESS &  --       & This paper\\
  1&59594.27--59594.40&   22&\ubvr&21686/21933& This paper\\
  1&60323.28--60323.41&   22&\ubvr&21686/21933& This paper\\
\noalign{\smallskip}\hline
\end{tabular}
\tablefoot{
In the column ``Station", the individual observing stations are
identified by the running numbers from the Praha/Zagreb photometric archives:
1 .. Hvar 0.65-m reflector, EMI tubes;
61 .. Hipparcos \hp\ photometry transformed into Johnson $V$ after \citet{hpvb};
110 .. TESS satellite broad-band photometry, sectors 31, 42, 43, 44, 70, and 71;
128 .. MOST satellite broadband photometry.
}
\end{table*}

\begin{table*}[h!]
\caption{
Dependent parameters of $\xi$~Tau models.
}
\centering  
\tiny
\begin{tabular}{llllllll}
                  &           & original & reference & {\bf all-data} & tides      & 5-component \\
var.              & unit      & val.     & val.      & val.           & val.       & val.        \\
\hline
\vrule height 8pt width 0pt
$m_1$             & $M_\odot$ & 2.232911 & 2.250038    & 2.270135    & 2.247897    & 2.241037    \\
$m_2$             & $M_\odot$ & 2.009948 & 2.133539    & 2.153717    & 2.131457    & 2.126487    \\
$m_3$             & $M_\odot$ & 3.734375 & 3.736787    & 3.781861    & 3.734669    & 3.828654    \\
$m_4$             & $M_\odot$ & 0.904359 & 1.427445    & 1.343642    & 1.442998    & 1.434199    \\
$m_5$             & $M_\odot$ &          &             &             &             & 0.779647    \\
$a_1$             & au        & 0.117567 & 0.118840    & 0.119203    & 0.118802    & 0.118693    \\
$a_2$             & au        & 1.082963 & 1.089639    & 1.093407    & 1.089315    & 1.092922    \\
$a_3$             & au        & 28.35343 & 29.43438    & 29.43633    & 29.44603    & 30.30589    \\
$a_4$             & au        &          &             &             &             & 0.020299    \\
$R_1$             & $R_\odot$ &          & 1.632       & 1.691       & 1.631       & 1.646       \\
$R_2$             & $R_\odot$ &          & 1.629       & 1.614       & 1.628       & 1.626       \\
$R_3$             & $R_\odot$ &          & 2.266       & 2.446       & 2.265       & 2.409       \\
$R_4$             & $R_\odot$ &          & 1.497$\dag$ & 1.424$\dag$ & 1.512$\dag$ & 1.500$\dag$ \\
$R_5$             & $R_\odot$ &          &             &             &             & 1.106$\dag$ \\
$P_{{\rm rot},1}$ & d         &          &             & 5.860       &             &             \\
$P_{{\rm rot},2}$ & d         &          &             & 6.148       &             &             \\
$P_{{\rm rot},3}$ & d         &          &             & 0.510       &             &             \\
$P_{{\rm rot},4}$ & d         &          &             & 1.126       &             &             \\
$P_{{\rm rot},5}$ & d         &          &             &             &             &             \\
$J_1$             & $\circ$   &          & 0.193       & 0.630       & 0.270       & 0.466       \\
$J_2$             & $\circ$   &          & 70.619      & 71.022      & 70.536      &             \\
\end{tabular}
\tablefoot{
The notation is as follows:
$m_i$ is the mass,
$a_i$ semimajor axis,
$R_i$ radius,
$P_{{\rm rot},i}$ rotation period,
$J_1$ mutual inclination between Aa+Ab and A+B orbits,
$J_2$ between A+B and AB+C orbits.
}
\label{dependent}
\end{table*}

\end{document}